\documentclass[twocolumn,apj]{emulateapj}
\usepackage{relsize,cancel, graphicx, dcolumn,graphicx,color,booktabs,bbold, microtype,braket,mathrsfs,mathtools,simplewick,tikz,empheq,amssymb,amsmath,bm}
\usepackage[colorlinks,plainpages=false,linkcolor=black,urlcolor=blue,citecolor=blue,pdfpagemode=UseNone,pdfstartview=FitH,allcolors=blue]
{hyperref}
\usepackage[T1]{fontenc}
\usepackage{cleveref}
\usepackage{amsmath}

\usepackage{tabularx}
\newcolumntype{L}{>{\raggedright\arraybackslash}X} 
\newcolumntype{C}{>{\centering\arraybackslash}X}   
\UseRawInputEncoding 

\begin{document}
\title{Exploring mutual information between IRIS spectral lines. I. Correlations between spectral lines during solar flares and within the quiet Sun.}
\author{Brandon Panos\altaffilmark{1,2}, Lucia Kleint\altaffilmark{1,2,3}, Sviatoslav Voloshynovskiy\altaffilmark{2}}\
\altaffiltext{1}{University of Applied Sciences and Arts Northwestern Switzerland, Bahnhofstrasse 6, 5210 Windisch, Switzerland}
\altaffiltext{2}{University of Geneva, CUI-SIP, 1205 Geneva, Switzerland}
\altaffiltext{3}{Leibniz-Institut f\"ur Sonnenphysik (KIS), Sch\"oneckstrasse 6, D-79104 Freiburg, Germany.}

\begin{abstract}
Spectral lines allow us to probe the thermodynamics of the solar atmosphere, but the shape of a single spectral line may be similar for different thermodynamic solutions. Multiline analyses are therefore crucial, but computationally cumbersome. We investigate correlations between several chromospheric and transition region lines to restrain the thermodynamic solutions of the solar atmosphere during flares. We used machine-learning methods to capture the statistical dependencies between 6 spectral lines sourced from 21 large solar flares observed by NASA's \textit{Interface Region Imaging Spectrograph} (IRIS). The techniques are based on an information-theoretic quantity called \textit{mutual information} (MI), which captures both linear and nonlinear correlations between spectral lines. The MI is estimated using both a categorical and numeric method, and performed separately for a collection of quiet Sun and flaring observations. Both approaches return consistent results, indicating weak correlations between spectral lines under quiet Sun conditions, and substantially enhanced correlations under flaring conditions, with some line-pairs such as \ion{Mg}{2} and \ion{C}{2} having a normalized MI score as high as $0.5$.  We find that certain spectral lines couple more readily than others, indicating a coherence in the solar atmosphere over many scale heights during flares, and that all line-pairs are correlated to the GOES derivative, indicating a positive relationship between correlation strength and energy input. Our methods provide a highly stable and flexible framework for quantifying dependencies between the physical quantities of the solar atmosphere, allowing us to obtain a three-dimensional picture of its state.
\end{abstract}


\section{Introduction}
The Sun emits radiation over the entire electromagnetic spectrum, with the bulk of its emission in the visible band. NASA's Interface Region Imaging Spectrograph (IRIS) small explorer spacecraft \citep{IRIS} captures UV radiation from plasma sensitive to temperatures in the range of 5000 to 10,000,000 K, corresponding to a height range between the photosphere and the corona. Part of this radiation is collected by IRIS's onboard spectrograph, which covers several strong chromospheric and transition region lines.

The shapes of these spectral lines are often related to the thermodynamics of the surrounding plasma, with each line's diagnostic capacity being determined via a forward modeling exercise of some quintessential model atom placed within an accurate solar atmosphere and allowed to radiate in accordance to a numerically feasible subset of radiative transfer constraints. Assuming accurate solar models, each line provides a different and complementary window of insight into the thermodynamic behavior of the solar atmosphere \citep[see for instance,][]{MgII_qs1, Lin_2015, CII_qs1}. 

The shape of a single spectral line, however, is often overly determined, and can be explained by a degenerate set of variations within the velocity fields, ionization ratios, densities and temperatures at a multitude of scale heights. This was clearly highlighted by \cite{Rubio_da_Costa_2017}, who  fed iteratively adjusted artificial flaring atmospheres (in an unconstrained way) from the radiative hydrodynamic code RADYN into the RH transfer code in an attempt to simulate the necessary thermodynamic conditions for the formation of a few interesting \ion{Mg}{2} h\&k flaring profiles. They found that single-peaked spectral shapes could be generated in a variety of ways: either by introducing large temperature and electron density spikes into the upper chromosphere, or by introducing unresolved up- and downflows within the same pixel. These results imply that the analysis of a single spectral line can easily be taken out of context. It is therefore crucial that one studies a multitude of lines simultaneously, leading to stronger physical constraints and less ambiguous conclusions.

An example of the utility of multiline studies can be found in a recent article by \cite{Chrom_Evap_CII_FeXXI}, who discovered several incompatibilities between observation and simulation by monitoring correlations between the Doppler shifts of both the \ion{C}{2} 1334.5 Å and \ion{Fe}{21} 1354.1 Å lines as a function of deposited heat flux for seven IRIS flare observations during their impulsive phases. Similarly, multiline inversion studies use the different formation heights of multiple spectral lines to derive atmospheric properties in three dimensions \citep[e.g.,][]{Vissers_2020,Vissers_2019,Solanki_2019,Johan_2019}.

The modern analysis of solar flare physics depends on a three-pillar self-referential system of checks, consisting of direct observations, simulations, and theoretical expectations. Any new results should be filtered through this system, with inconsistencies warranting either changes to our current understanding of the standard flare model \citep[e.g.][]{standard_model2}, or to the underlying assumptions of our simulations. The above-mentioned studies call for the use of multiline observations to challenge our current models and restrain the parameter space of solutions.

These considerations naturally promote the question about interline correlations, a topic that has not received enough attention in the solar physics community given the weight of its importance both as a diagnostic tool and as a way of  providing strict multi-height benchmarks for current state-of-the-art hydrodynamic and radiative transfer codes \citep[e.g.,][]{Testa_2020,Graham_2020}. Although there have been a few such studies, see \cite{correlation_overview} for a multiwavelength spectroscopic overview, the analyses seldom deal with spectral shapes and rely almost exclusively on simple linear correlation based metrics such as covariance to capture dependencies.

The objective of this paper is to provide a machine-learning based model derived from 21 large solar flares, that can highlight the dependencies between a multitude of chromospheric and transition region lines observed by IRIS. These dependencies or correlation can then be used to understand the physics of the solar atmosphere at various heights. The spectral line shapes compared are from the ions: \ion{Mg}{2}, \ion{C}{2}, \ion{Si}{4}, \ion{O}{4}, \ion{Fe}{2} and \ion{Fe}{21}. The final model will calculate the degree of correlation between any two of the above mentioned spectral lines. To do this in the most rigorous way possible, we make use of an information-theoretic quantity called the mutual information (MI), which is capable of picking up both linear and nonlinear dependencies \citep{MI_vs_CORR}.\\ 

This paper is organized as follows: In Section \ref{Data_preprocessing}, we address the preprocessing steps necessary to prepare the data for our particular research goals. In particular, we use a \textit{variational autoencoder} (VAE) to generate a clean flaring data set. In section \ref{MI_section}, we discuss the concept of mutual information, and provide both a categorical and numerical method for estimating it. The relevant intuitions and mathematical concepts are addressed and the convergence properties of the numerical scheme are subsequently investigated. In section \ref{Results_section}, we apply the above technology to both the quiet Sun and flaring data sets, and analyze the evolution of the latter with respect to the GOES curve and derivative of a single flare. We also investigate the changes in correlation strengths when considering spectra only from the flare ribbon. In section \ref{Discussion_section}, we discuss some of the result, including the meaning of low correlation scores, and finally conclude and offer potential attractive extensions and research avenues in section \ref{conclusion_section}. 

\begin{table}[t]
\caption{Quiet Sun observations}\centering 
\begin{tabularx}{.45\textwidth} { >{\hsize=.1cm\raggedright\arraybackslash}X >{\centering\arraybackslash}X >{\centering\arraybackslash}X>{\centering\arraybackslash}X>{\raggedleft\arraybackslash}X }
\toprule\toprule
\#&Date&Time Obs Start&OBSID\\ 
\midrule
1& 2014-02-06&12:44&3803257203\\
2&2014-02-07&11:29&3803257203 \\
3&2014-02-08&13:32&3803257203\\
4&2014-02-11&05:10&3864255653\\
5&2014-05-11&06:49&3800258458\\
6&2014-05-15&14:09&3820357403\\
7&2014-05-16&07:58&3800258458 \\
8&2014-06-26&22:32&3820005183\\
9&2014-06-28&16:48&3820505482\\
10&2014-06-28&21:42&3820505482\\
11&2014-09-02&07:30&3800258465\\
12&2014-10-07&07:54&3800258458\\
13&2014-12-14&15:38&3800106080\\
14&2014-12-14&17:15&3800106080\\
15&2014-02-10&12:39&3803257203\\
16&2014-09-25&22:09&3820107266\\
17&2015-02-18&20:01&3800010066\\
18&2015-10-10&23:34&3620005935\\
\bottomrule
\label{obs_qs}
\end{tabularx}
\end{table}
\begin{table}[h]
\caption{Flare observations}\centering 
\begin{tabularx}{.45\textwidth} { >{\hsize=.1cm\raggedright\arraybackslash}X >{\centering\arraybackslash}X >{\centering\arraybackslash}X>{\centering\arraybackslash}X>{\raggedleft\arraybackslash}X }
\toprule\toprule
\#& Class&Date&Time Obs Start&OBSID\\ 
\midrule
1&M1.0& 2014-06-12&11:09&3863605329\\
2&M1.0&2014-11-07&09:37&3860602088 \\
3&M1.1&2014-06-12&18:44&3863605329\\
4&M1.1&2014-09-06&11:23&3820259253\\
5&M1.1&2015-08-21&16:01&3660104044\\
6&M1.3&2014-10-26&18:52&3864111353\\
7&M1.4&2015-03-12&05:45&3860107053 \\
8&M1.8&2015-03-11&04:46&3860259280\\
9&M2.3&2014-11-09&15:17&3860258971\\
10&M2.9&2015-08-27&05:37&3860605380\\
11&M3.4&2014-10-27&20:56&3864111353\\
12&M3.9&2014-06-11&18:19&3863605329\\
13&M6.5&2015-06-22&17:00&3660100039\\
14&M7.3&2014-04-18&12:33&3820259153\\
15&M8.7&2014-10-21&18:10&3860261353\\
16&X1.0&2014-03-29&14:09&3860258481\\
17&X1.6&2014-09-10&11:28&3860259453\\
18&X1.6&2014-10-22&08:18&3860261381\\
19&X2.0&2014-10-27&14:04&3860354980\\
20&X2.1&2015-03-11&15:19&3860107071\\
21&X3.1&2014-10-24&20:52&3860111353\\
\bottomrule
\label{obs}
\end{tabularx}
\end{table}

\begin{figure*}[t] 
\centering
\includegraphics[width=1\textwidth]{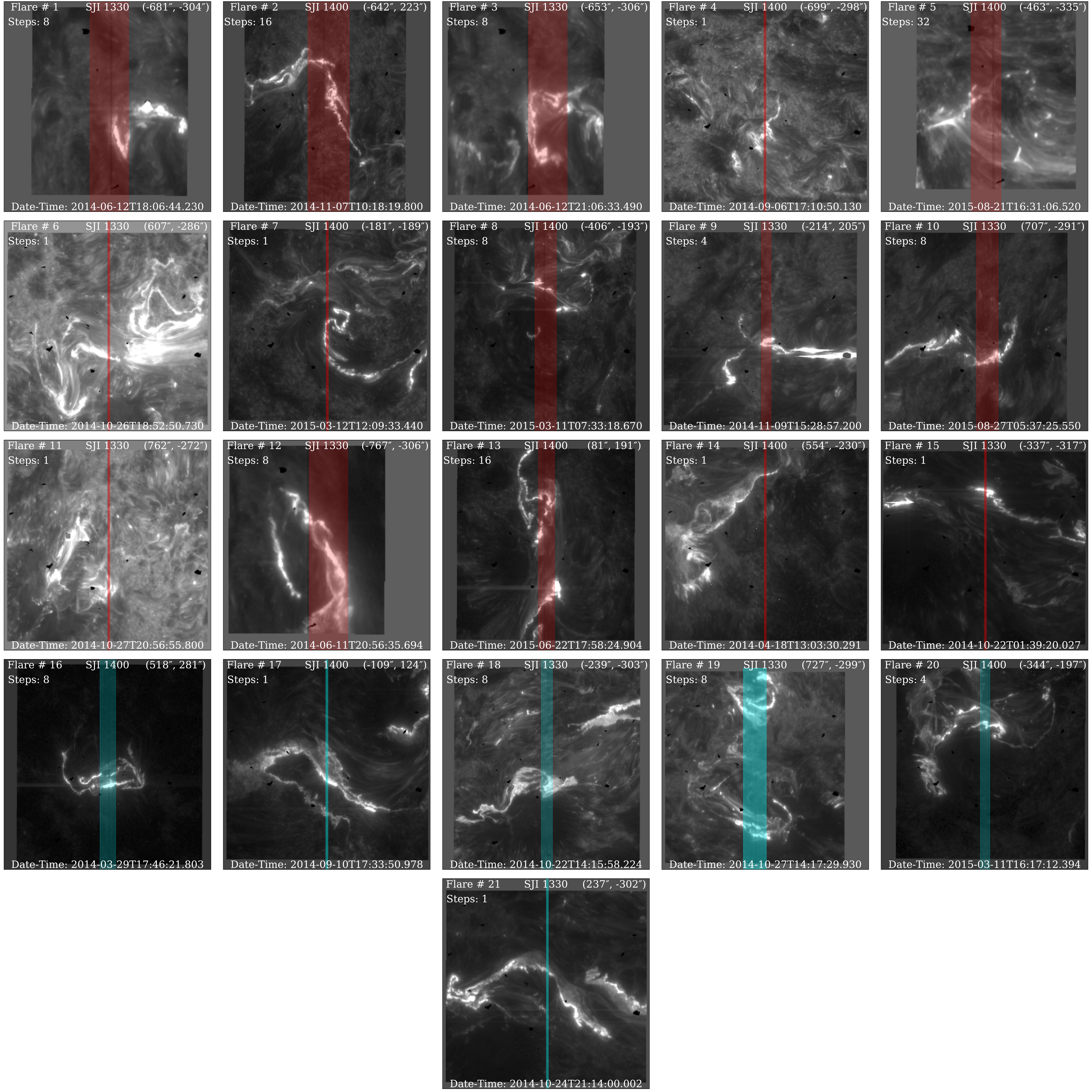}
\caption{Collage of IRIS slit-jaw-images (SJI) showing a single snapshot from each of the 21 flares listed in \Cref{obs}. Spectral data was collected from the highlighted regions, with red and cyan indicating which flares are M- and X-class, respectively. These particular flares were chosen on account of the good alignment between IRIS' spectrometer and the flaring activity seen in white. Note that spectra were only collected from evenly spaced steps within the colored regions, such that the spectrometer need not cover the entire colored region. The number of steps is indicated on the left-hand side of every image, along with the flare number on top, SJI filter, central position of the image in arcseconds, and the time each individual SJI was rendered at the base of the image.}
\label{sji_collage}
\end{figure*}

\section{Data}
\label{Data_preprocessing}
The assumptions underlying line formation in the quiet Sun are often not extendable to flaring atmospheres, with key differences underpinning the radiative transfer in both cases. The additional energy supplied by the electron beam during a solar flare can understandably lead to an increase in line correlation by either reducing the thermalization lengths and thereby forcing a type of "locality" (see for example a study by \cite{Kerr_2019} where the \ion{Si}{4} line formation becomes optically thick), or by inducing bulk velocity flows which could knit the atmosphere together at multiple scale heights. In anticipation of these differences, we chose to analyze line correlations in the quiet Sun and flaring atmospheres separately. We therefore created two data sets consisting of 18 quiet Sun observations and 21 flare observations sourced from the years 2014 and 2015. All observations were on disk and encompassed a rich variety of modes, including sit-and-stares, rasters with both fast and slow cadence rates, and a few observations with roll angles offset from solar north. A list of the quiet Sun observations can be found in Table \ref{obs_qs}, while the flare observations can be found in Table \ref{obs}, along with an accompanying collage of SJI's in \Cref{sji_collage} taken at a critical moment in each flare's development. We restricted our analysis to large M- and X-class flares, with the IRIS slit positioned directly over the flare ribbon. Each IRIS observation also covers all six of the above mentioned spectral windows, coming to a total of $4,788,392\times 6 $ spectra for flare observations, and $2,977,090 \times 6$ spectra for quiet Sun observations.\\

In order to make the data compatible with our particular research goals and methods, an additional layer of preprocessing was required on top of the IRIS level 2 data product. Spectra with missing data in the form of large negative values as well as overexposed spectra with more than five consecutive points at the same intensity were removed. The spectra associated with each line were then interpolated onto the same wavelength grid, such that different observations always contained the same number of wavelength ($\lambda$) points. Note that we only require the number of bins to be consistent within each spectral line, for instance, all \ion{Mg}{2} spectra should have the same number of $\lambda$ points regardless of the observation, while \ion{Si}{4} spectra can have its own self consistent $\lambda$ grid. This step is important for the machine-learning methods used to calculate the line correlations. Finally, the intensity degree of freedom was removed from the data by normalizing each spectrum by its maximum value, such that we only study different line shapes.

\subsection{Separating quiet Sun and flaring spectra}
\label{VAE_sec}
The spectrograph of IRIS can scan a field of view (FOV) of $130\times175~\text{arcsec}^2$, and therefore most flare observations in Table \ref{obs} also contain a number of quiet Sun spectra toward the observation's peripheries. In addition to this, a region might be both flaring and not flaring at different instances in time. Since our research methodology relies on the generation of two clean data sets containing either quiet Sun or flaring spectra exclusively, we had to find a reliable way of isolating the flaring region of each observation. The criterion for determining the flaring region was based on characteristics derived from a single spectral line. For reasons that will be made clear at the end of this section, we selected the \ion{Mg}{2} h\&k lines. As a first approximation, we could introduce an intensity threshold and simply discard any spectrum that falls below this threshold. This approach is well founded since strong lines typically experience an increase of  2-3 orders of magnitude in intensity during solar flares \citep{Kerr_2015}; however, there are several noticeable deficiencies with this approach. The selection of a single threshold is not a trivial task, and experience indicates that a threshold that isolates a flaring region appropriately for one observation does not necessarily translate well to a different observation. In addition to this, a simple threshold method may discard interesting low-intensity flare spectra.

An alternative and, in our opinion, more forgiving approach is to isolate flaring regions based on spectral shape. \citet{Panos_2020} showed that \ion{Mg}{2} h\&k line profiles associated with quiet Sun and flaring atmospheres could easily be distinguished using neural networks (NNs). Although their models incorporated an intensity component, features such as k/h ratio and subordinate line emission appeared to be strongly descriptive of the two classes. 

\begin{figure}[t] 
\centering
\includegraphics[width=.5\textwidth]{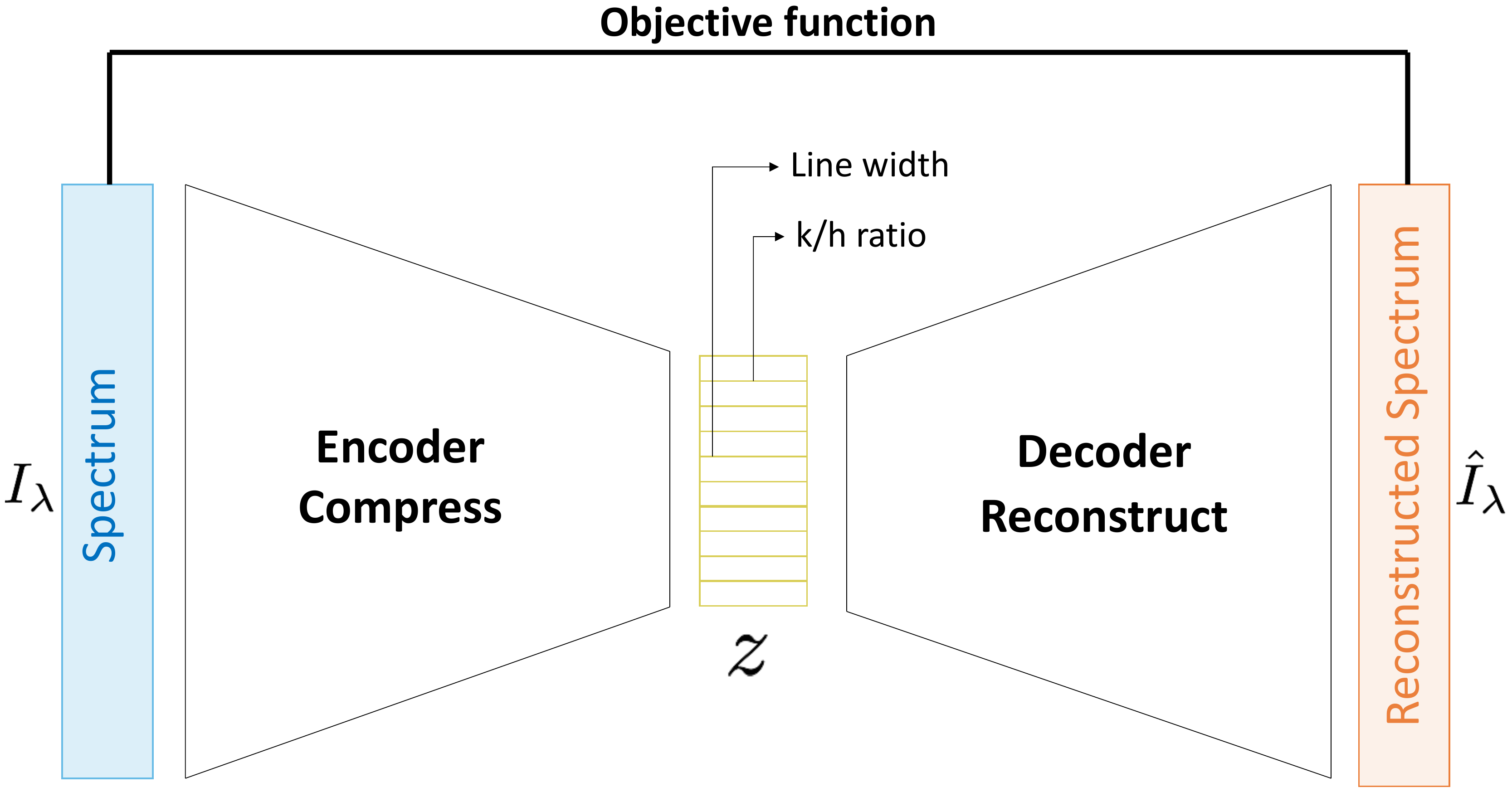} 
\caption{VAE architecture trained on $\sim 3$ million quiet Sun \ion{Mg}{2} spectra from 18 different observations. The VAE takes spectra, compresses them into a 10-dimensional latent space $z$, (here for example indicated by line width and k/h ratio), and then tries to reconstruct the spectra such that they look as similar to the originals as possible. The difference between the original and reconstruction is called the reconstruction error, which is substantially larger for flare-type spectra (unseen at training time). The reconstruction error can be used to reliably and consistently isolate flaring regions.}
\label{VAE_arch}
\end{figure}

\begin{figure*}[tbh]
\includegraphics[width=1\textwidth]{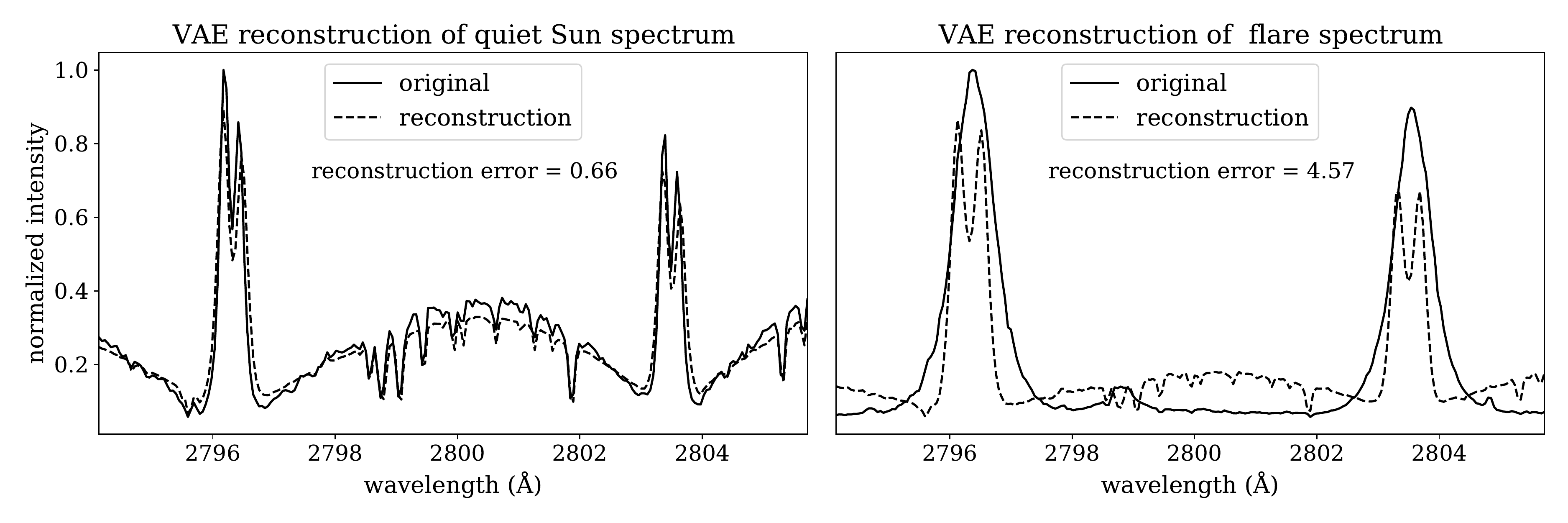}
\caption{ Reconstruction of spectral profiles from a VAE trained on 3 million quiet Sun spectra over 18 different IRIS observations. The left panel shows a comparison between the original (solid line) and reconstructed (dashed line) quiet Sun spectrum. The difference between the two spectra is referred to as the reconstruction error, characterized by $\sum_i||I_{\lambda_i}-\hat{I}_{\lambda_i}||^2_L$. Since the VAE has been trained to encode and decode quiet Sun spectra, it is much more competent at generating realistic quiet Sun profiles, and therefore has a relatively low reconstruction error in comparison to the error associated with the reconstruction of flaring profiles, as seen in the right panel.}
\label{reconstruction}
\end{figure*}
\begin{figure*}[tbh]
\centering
\includegraphics[width=.48\textwidth]{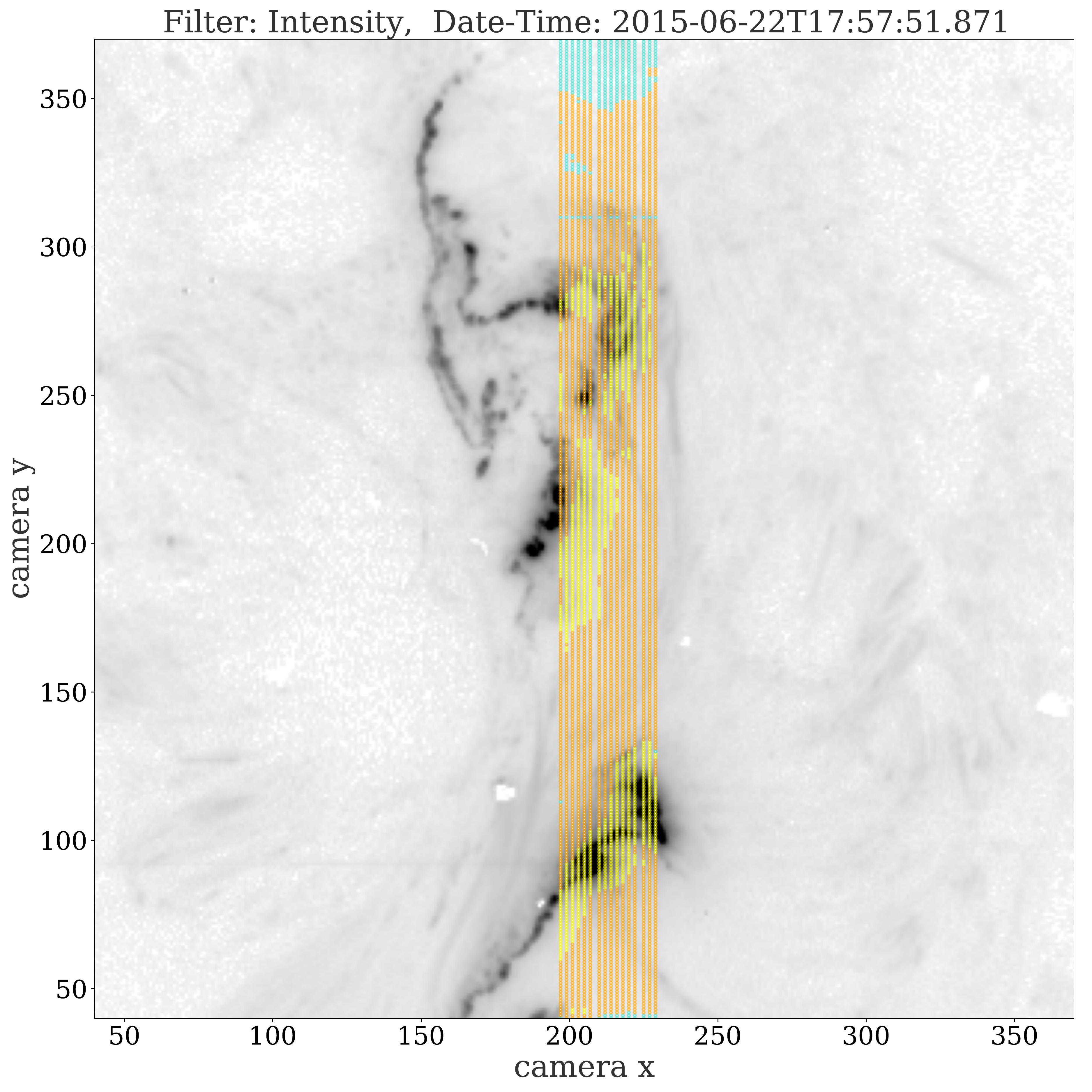}
\includegraphics[width=.48\textwidth]{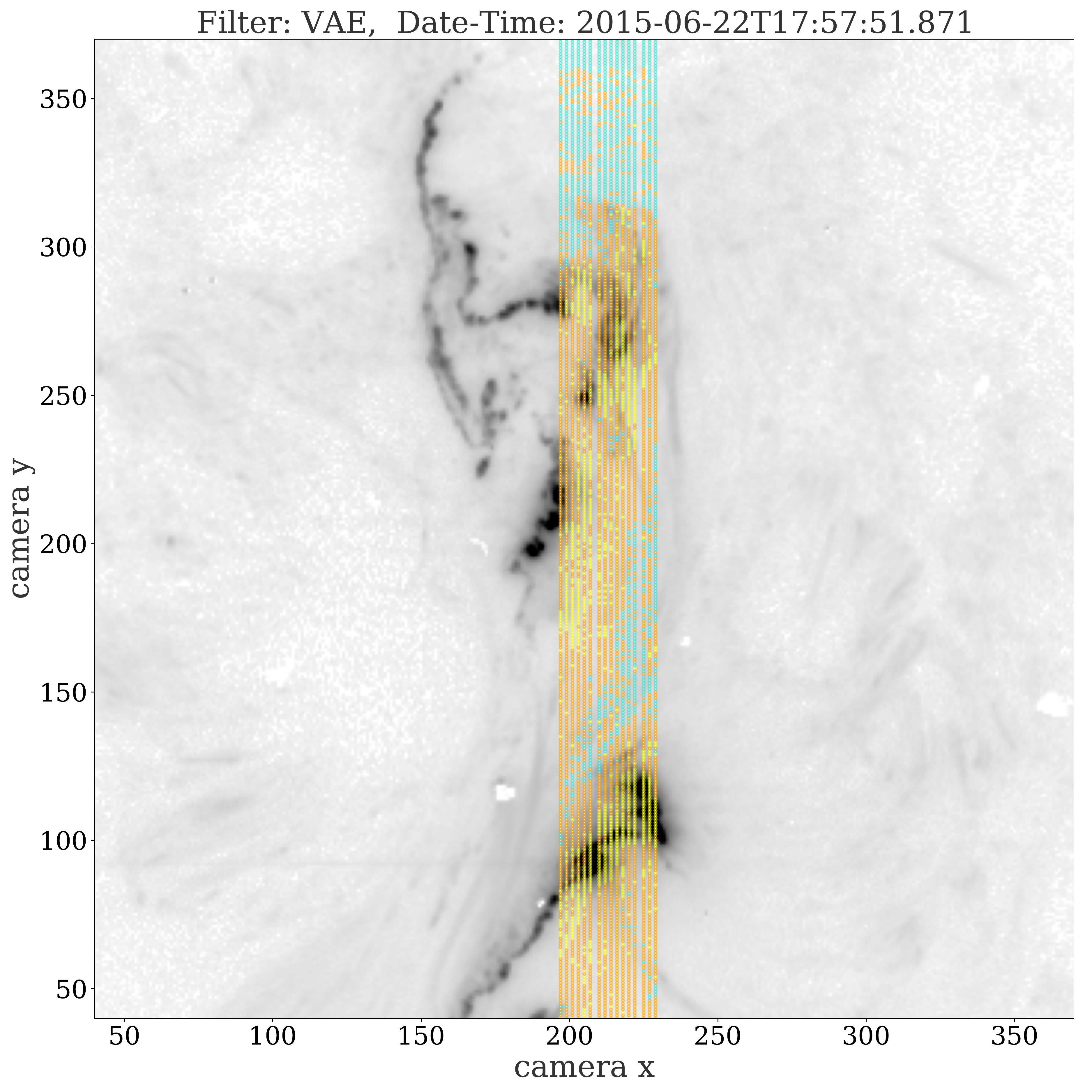}
\includegraphics[width=.48\textwidth]{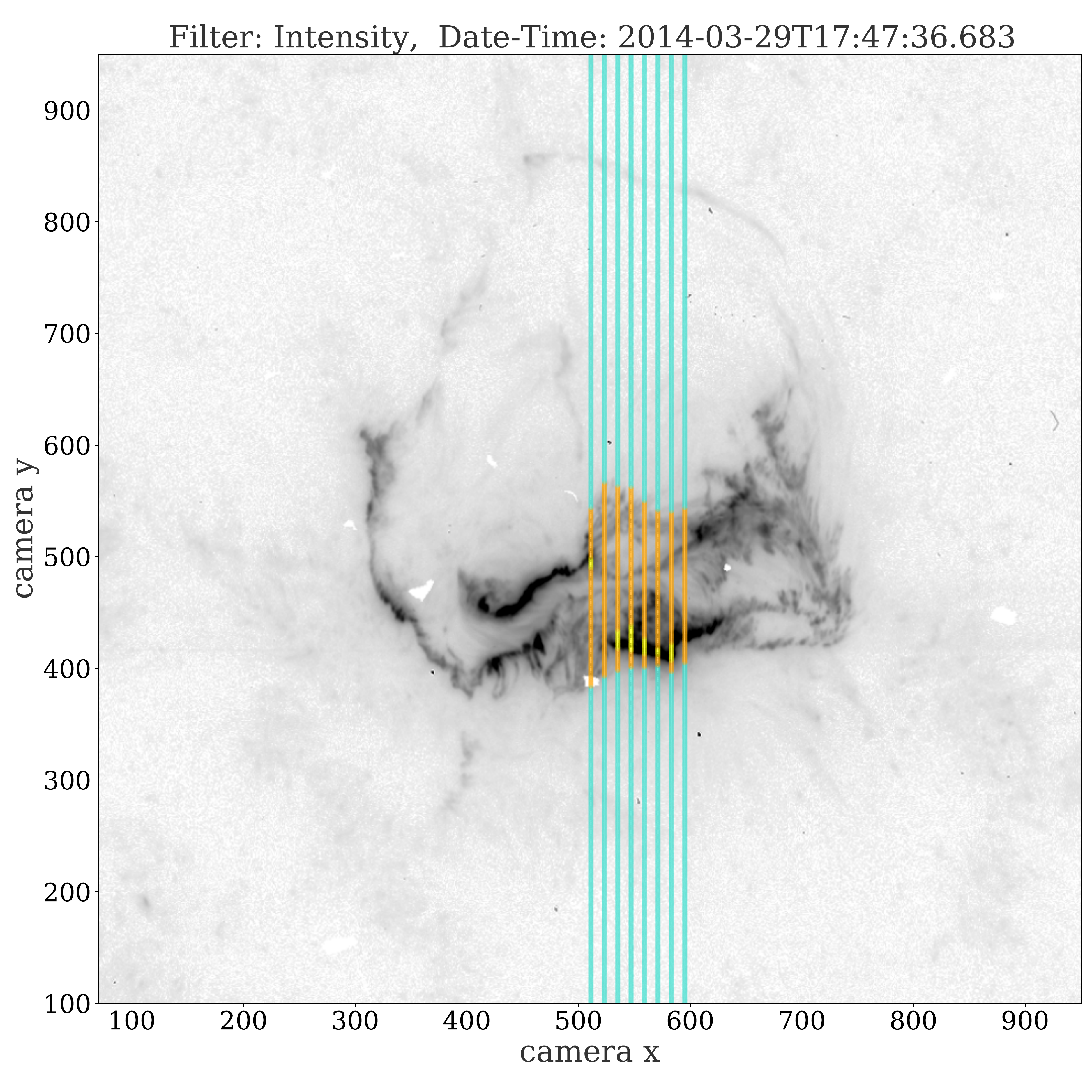}
\includegraphics[width=.48\textwidth]{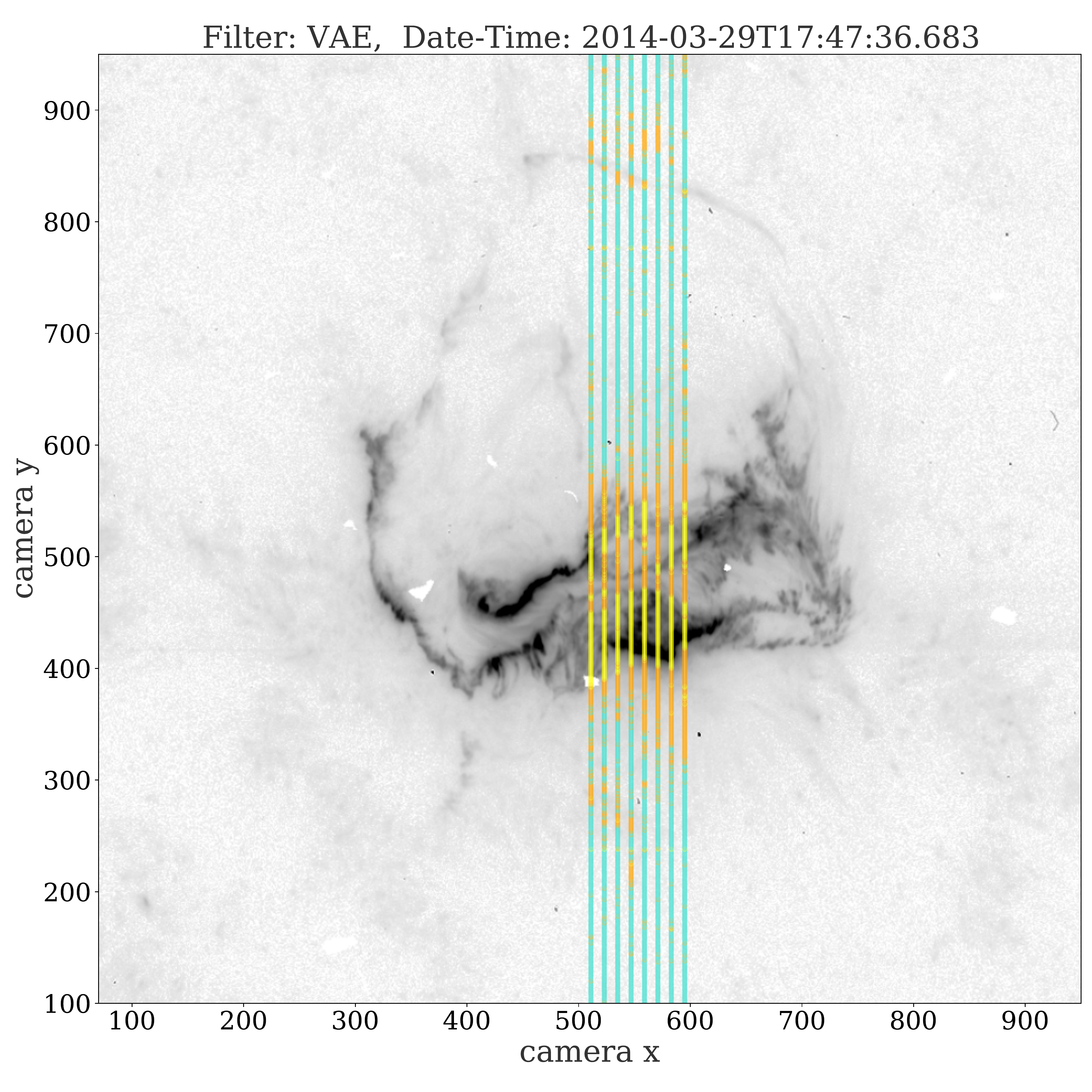}
\caption{Comparison between intensity and VAE masks for two different solar flares based on the \ion{Mg}{2} line. The intensity masks (first column), group spectra in bins of $\text{Blue}\in [0,200)$,  $\text{Orange}\in [200,1000)$ and $\text{Yellow}\in [1000,\infty)~\text{DN/s}$. In contrast, The VAE masks (right column) group spectra in terms of reconstruction errors 0.5 (Blue), 3 (Orange) and 5 (Yellow). In both cases, there is an expectation for higher intensities and reconstruction errors to better isolate the flaring regions. It is often the case that VAE maps are more consistent and generalize better than their intensity counterparts. For instance, both the intensity and VAE schemes map the ribbon of the first flare to Yellow, however, only the VAE under the same parameter settings manages to map the ribbon of the second flare in a consistent way. Additionally, low-intensity spectra which are clearly flare induced, such as those appearing in the second flare around the $y=850$ pixel mark, are consistently captured by the VAE, and missed by the intensity mask.}
\label{qs_masks}
\end{figure*}

In line with this research, we implemented a dual NN architecture, VAE, to learn the salient features associated with quiet Sun spectral shapes; see Figure \ref{VAE_arch}. A VAE is a type of NN that learns how to compress/encode and reconstruct/decode data. For instance, a VAE can take a \ion{Mg}{2} spectrum consisting of 240 $\lambda$ points and compress it down into a \textit{latent space} representation ($z$) of just a few numbers (in our case 10). These number could represent things such as the k/h ratio, line width, or integrated emission between the line cores; however, it is up to the network to decide what the most efficient basis is for representing the spectra. The VAE then tries to learn a set of instructions that will allow it to unfold the efficient 10-dimensional representation back into a full 240-dimensional vector, with the objective of making the reconstructed spectrum as similar to the original as possible. The function that guides the learning process of the network is called an objective function. In this case the objective function is the mean squared difference between each $\lambda$ point's intensity $\sum_i||I_{\lambda_i}-\hat{I}_{\lambda_i}||^2_L$, where $I_\lambda$ is the original and $\hat{I}_\lambda$ is the reconstructed spectrum, with $i\in[1\cdots,240], ~I_\lambda\in\mathbb{R}^{240}$. The VAE adjusts its parameters via a process called backpropagation, which allows the network to efficiently minimize the objective function via a simple gradient descent method. If the difference between the reconstruction and original (called the reconstruction error) is small, then the network has learned: 1) the most important features of the spectrum, and 2) a set of instructions for rebuilding the spectrum from these features. In essence, a VAE attempts to learn the relationships and dependencies between the intensity values at each $\lambda$ point of the spectrum, by looking at how they are connected over the entire data set. As an example, the quasi-continuum emission between the \ion{Mg}{2} line cores always has a parabolic shape: sometimes more flat, sometimes with a higher continuum value between the two spectral lines. It is therefore inefficient to store all intensity values at each of the $\lambda$ points, when a single number representing the integrated emission could provide a good estimate of the shape of this continuum region.

The learned relationships and dependencies between the intensity values at each $\lambda$ point depend on the type of data the network is trained on. In machine-learning terminology, the VAE learns a model for the particular training distribution (in our case, quiet Sun spectra). The relationships between the intensities at each $\lambda$ point may follow an entirely different set of rules for a distribution of flaring spectra. If we therefore train the VAE only on quiet Sun spectra, it will not have the vocabulary necessary to compress and reconstruct a spectrum from a solar flare, without incurring a comparatively large reconstruction error. An example of this can be seen in Figure \ref{reconstruction}. In the left panel, the reconstructed spectrum (dashed line) is almost identical to the original quiet Sun spectrum (solid line); however, the flaring spectrum on the right is very poorly approximated by its reconstructed counterpart. Notice that although the approximation is not appropriate, the VAE nevertheless attempts to create a profile that has broad cores and a flat quasi-continuum. Since it does not know how to construct something outside of the set of rules governing the quiet Sun, it cannot form a single peak. This reconstruction feature of a VAE is useful for solving our problem. We therefore trained a VAE on the entire quiet Sun data set of 18 IRIS observations, and used the reconstruction error to isolate the flaring regions. The regions with low reconstruction error were not used in any subsequent analysis.

A comparison between the intensity threshold and VAE method for the \ion{Mg}{2} line can be seen in Figure \ref{qs_masks}. We have displayed the results for two different solar flares. In both cases, there is an expectation for higher intensities and reconstruction errors $(\text{blue}< \text{orange}<\text{yellow})$ to better isolate the flaring regions; however, the affiliation of higher intensities and reconstruction errors with more flare-like spectra is somewhat arbitrary. For instance, a low-intensity spectrum might have a particular shape that never occurs within the quiet Sun. Under these circumstances, the spectrum is as flare-like as possible, nevertheless, the spectrum will be assigned to a lower-ranked bin (orange) under the intensity framework.

The definition of quiet Sun and flare from the limited perspective of a single spectral line is again somewhat arbitrary. For instance, \ion{Fe}{2} only deviates in shape substantially from its quiet Sun counterpart directly over the flare ribbon. This leads to a nested set of quiet Sun masks (one for each spectral line). Selecting the strongest criteria where all lines agree upon the flaring region would result in a stunted analysis of only the most intense parts of solar flares. For this reason, we chose to define the flaring region as the combination of yellow and orange pixels from the perspective of the \ion{Mg}{2} line, which most often encompassed all other masks. \\

We have simplified our description of the VAE architecture for clarity, the full mathematical background can be found in \citet{vae_tut}. For the most part we have described a simpler model called an autoencoder. The difference between an ordinary autoencoder and a VAE can be found in the smoothness of the latent space $z$. Autoencoders do not respect the shape of the latent space, and are allowed to contort it pathologically to meet the needs of the objective function. On the contrary, a VAE has an additional term in the objective function that not only tries to minimize the difference between the original and reconstructed spectra, but also tries to promote the construction of a smooth latent space. On a high level, this means a VAE can interpolate to new latent space points, and generate spectra that have never been seen before, but are still convincingly quiet Sun like. This particular quality of a VAE elevates it into a class of models called generative models. The generative nature of VAEs make them especially appropriate for our purposes, since in theory, any quiet Sun spectrum that we missed within our 18 observations, can be constructed by the VAE and still assigned a small reconstruction error.

 \begin{figure}[t] 
\includegraphics[width=.235\textwidth]{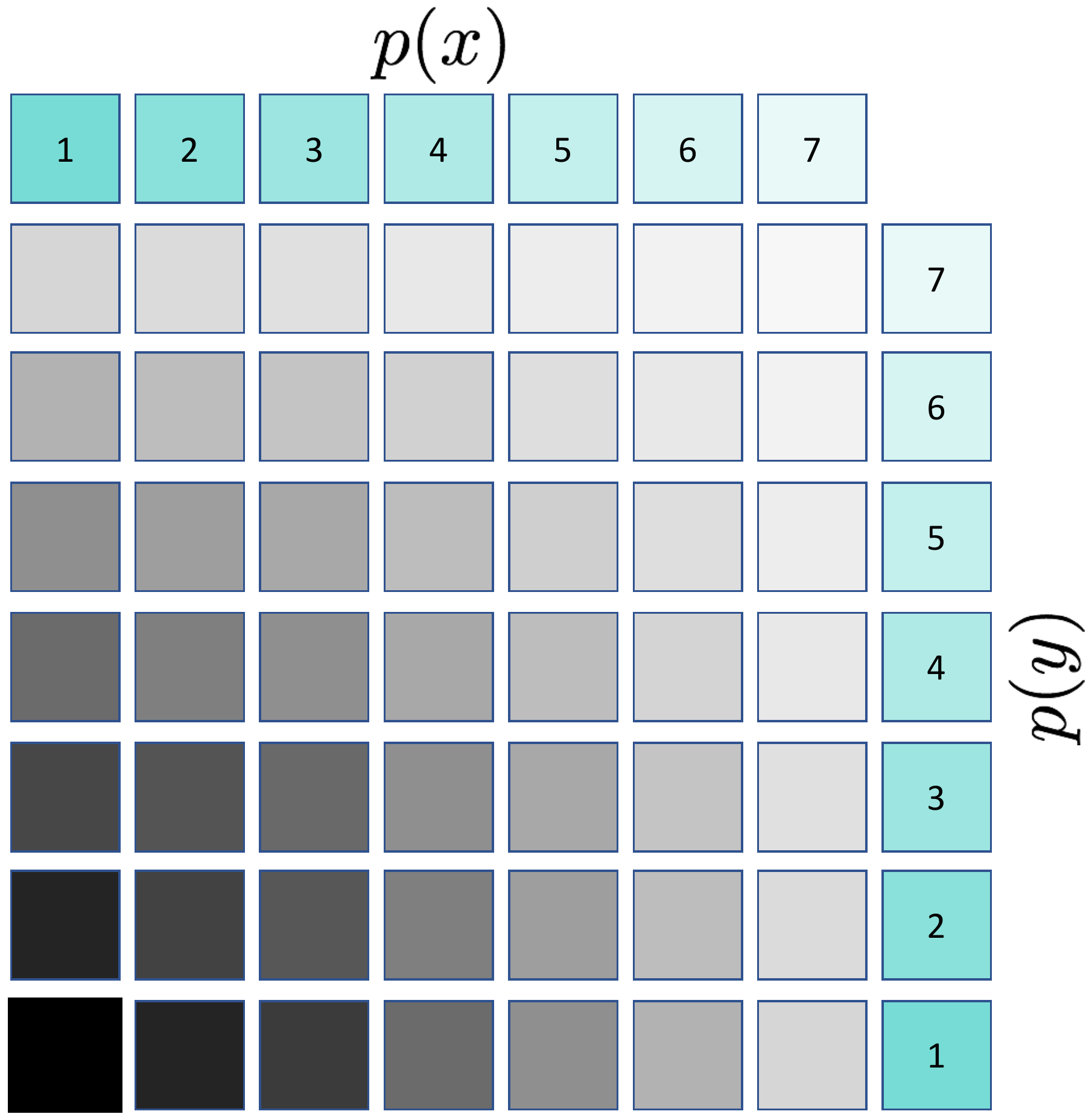} 
\includegraphics[width=.235\textwidth]{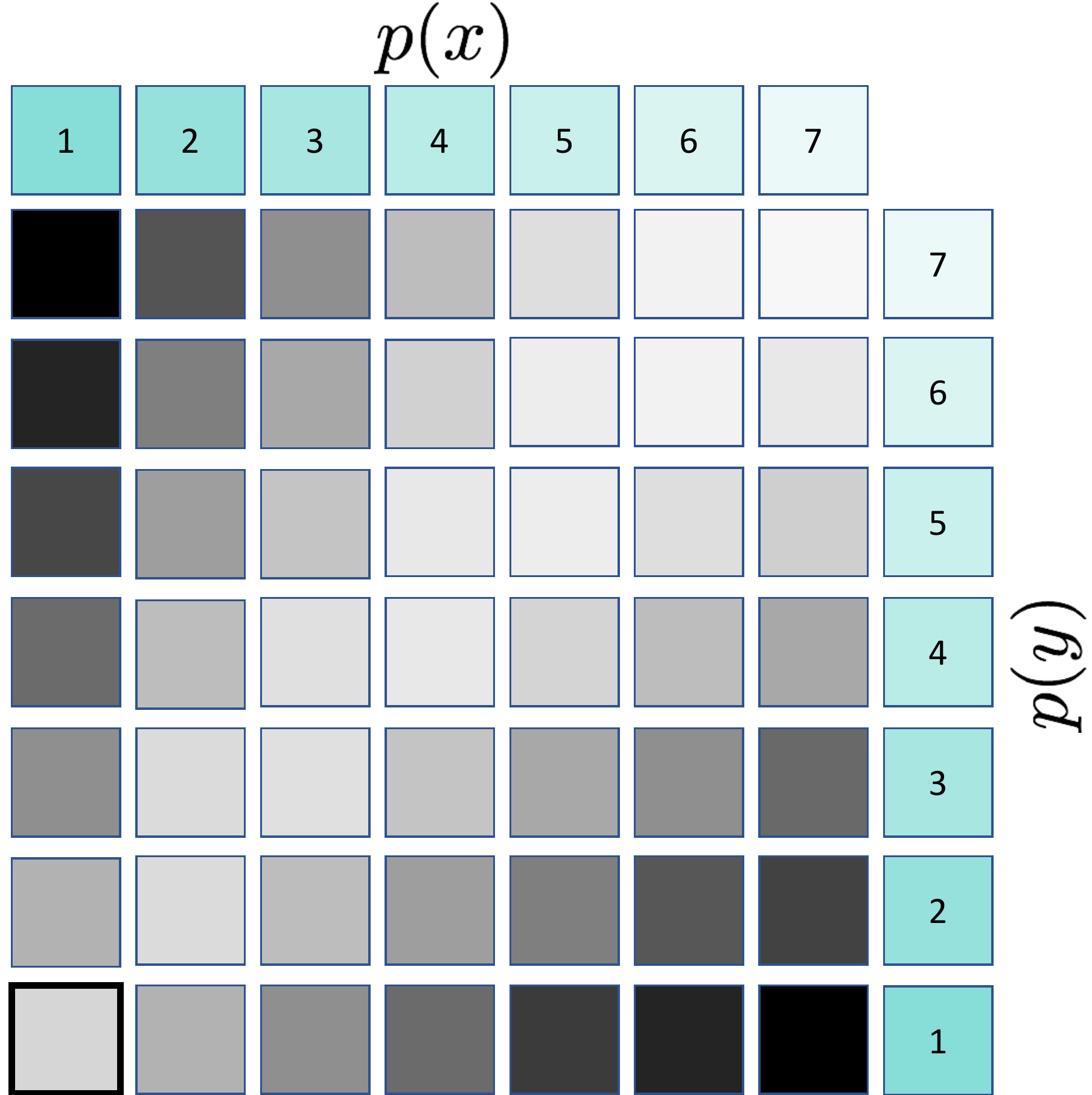} 
\caption{An example of an independent and a true joint probability distribution (left and right panels, respectively). The marginal distributions $p(x)$ and $p(y)$ in blue represent the probability of observing specific spectral shapes falling within the 7 shape categories of line $\mathcal{L}1$ and $\mathcal{L}2$ respectively. For instance, $p(x_1)$ may represent the probability of observing a single-peaked \ion{Mg}{2} spectral shape, while $p(y_1)$ could correspond to the probability of observing blue Doppler shifted \ion{Si}{4} shapes. If the shapes from one line are completely independent from the shapes in the other line, the joint probability $p(x,y)$ (left grid) reflects the frequencies of the marginal distributions, which for our example would mean that \ion{Mg}{2} spectra with single peaks, would occur most often with blueshifted \ion{Si}{4} spectra, purely based on their occurrence frequencies and not necessarily any physical correlation. On the other hand, if the joint probability decouples from the marginal distributions (right panel), there exists a correlation between the two spectral lines. In our fictitious example, it may mean that single-peaked Mg spectra $x_1$, are correlated to broad \ion{Si}{4} spectra $y_7$, but rarely occur together with \ion{Si}{4} blueshifts $y_1$, even though blueshifts are very common in \ion{Si}{4}. The degree of this correlation can be calculated as the difference between the left and right distributions, which is the MI.}
\label{prob}
\end{figure}

\section{Calculating the correlation between spectral lines}
\label{MI_section}
In this section, we develop the intuition and notation necessary to understand an information-theoretic quantity that will allow us to calculate the correlations between lines in a robust way. We then provide two different approaches for estimating the correlation between pairs of spectral lines, both under quiet Sun and flaring conditions. 

\subsection{Mutual Information}
In probability theory, a random variable $X$ is associated with a number of possible states $x_i\in X$. The marginal probability distribution $p(x)$ tells us how likely each of theses state are. Similarly, we can have another random variable $Y$ described by the marginal distribution $p(y)$. If the two random variables are completely independent, then the probability $p(x_i,y_j)$ of simultaneously observing a particular pair of states, called the joint probability, would be given by the product of the probabilities of observing them individually $p(x_i)p(y_j)$. This is commonly referred to as the product of the marginals, and is precisely the behavior we would expect from two fair dice throws. The concept of variable independence is depicted graphically in the left panel of Figure \ref{prob}, with the blue horizontal and vertical blocks corresponding to the marginal probability distributions of $X$ and $Y$ respectively, whose products give the resulting joint probability distribution $p(x,y)$ in black. We have allowed both random variables to have 7 degrees of freedom (states), and assigned darker shades to higher probabilities. This means that this example is not similar to a dice throw, because we arbitrarily selected the probability of each state, with state 1 being more likely than e.g. state 2. For the example, we selected the same probabilities for $X$ and $Y$. As a consequence of the high frequency of state 1, $p(x_1,y_1)$ in the lower left corner is predicted to occur most often. The fact that these pairs occur with a particularly high frequency, has nothing to do with any physical relationship between them, but is merely the byproduct of a statistical rule.  In conclusion, the defining quality of two uncorrelated variables is that the pattern of the joint probability distribution mirrors the frequencies of the marginals. If we apply this to the example of two fair dice throws, the left panel of Figure \ref{prob} would have an equal grey color everywhere because any combination of two dices is equally likely.

On the other hand, the panel on the right of Figure \ref{prob} shows a distinct decoupling between the marginals and joint distributions, with the probability of $p(x_1,y_1)$ this time being small despite the high frequencies associated with states 1. The departure from the rules governing independent variable behavior naturally implies a degree of dependence between the variables. This affords us a mechanism for calculating the correlation between two random variables as the degree to which the black distributions in the left- and right-hand panels of Figure \ref{prob} deviate from one another. We refer to these distributions as the independent $P^\dagger$ and true $P$ probability distributions respectively. An example is described in Fig.~\ref{prob}.

The Kullback-Leibler or $\mathrm{KL}$-divergence is commonly used as a statistical measure of distance between two probability distributions. We will use this measure to calculate the degree to which $P^\dagger$ and $P$ disagree. In general, the $\mathrm{KL}$-divergence between two distributions $p$ and $q$ is defined as:
\begin{equation}
\begin{aligned}
 D_\mathrm{K L}(p \| q) &=\sum_x p(x) \log \frac{p(x)}{q(x)}, \\ 
 &=\mathbb{E}_{p(x)} \left[\log \frac{p(x)}{q(x)}\right],
 \label{KL}
\end{aligned}
\end{equation}
where $E_p$ is the expectation value of the log difference between both distributions with respect to $p(x)$. Notice that the distance between both distributions is simply the sum of the logarithmic difference between the probabilities at each point, i.e., we superimpose the left and right-hand panels of Figure \ref{prob} and take the log difference between the shades. The prefactor of  $p(x)$ makes the $\mathrm{KL}$-divergence asymmetric and not a metric in the common sense (since $D_\mathrm{K L}(p \| q) \neq D_\mathrm{K L}(q \| p) $, and the triangle inequality is not necessarily respected), however the addition of this factor along with the logarithmic terms encourage interpretations within the field of information theory, where the asymmetry is a desirable quality rather than a glitch. We suggest the text of \cite{infomation} for readers interested in an information-theoretic understanding of this quantity. The dependence between two random variables is then simply the distance between the true and independent distributions $D_\mathrm{K L}(P \| P^\dagger) $, which is called the mutual information and is explicitly written as: 
\begin{equation}
MI(X ; Y) = \sum_x\sum_y p(x, y) \log \frac{p(x, y)}{p(x) p(y)}.
\label{mutual_information}
\end{equation}
Unlike covariance, MI can capture nonlinear dependencies and is amenable to categorical data, where the symbols used to represent each category do not have a meaningful mapping into $\mathcal{R}^N$ \citep{MI_2014}. In this sense, MI is a natural extension of covariance, and is a complete measure of all possible dependencies. 

It is simple to calculate Eq.~\ref{mutual_information} if both $p(x,y)$ and the marginal distributions are known, but in practise, we only have access to data generated from the joint distribution. Thus, the challenge is to estimate the MI from data under unknown $p(x)$ and $p(y)$. We will now look at two different approaches for estimating the MI. The first approach tries to derive $P$ and $P^\dagger$ directly, while the second infers the distributions from their effects on a third test distribution.

\subsection{Approach 1: Categorical}
\label{cat_sec}
At face value, the procedure for calculating the MI between spectral lines appears relatively simple: count the co-occurrences of states, construct the true and independent distributions, and finally compute the MI using the $\mathrm{KL}$-divergence as a way to quantify the distance between these distributions. However, there are a number of practical nuances when applying this procedure to spectral lines.

In the information-theoretic frame, each spectral line $\mathcal{L}i$ represents a random variable. The states of these random variables correspond to the different spectral shapes each specific line can obtain. For instance, if $\mathcal{L}1$ corresponds to the \ion{Mg}{2} line, then states $x_1$ and $x_2$ could represent spectra that have single-peaked shapes or deep central reversal respectively.  By counting the number of co-occurrences between the states of two spectral lines $\mathcal{L}1$ and $\mathcal{L}2$ within the same pixel, we sample from the joint distribution $p(x,y)$, and can construct the true probability distribution $P$, where the value of element $P_{ij}$ is proportional to the number of times state $x_i$ from $\mathcal{L}1$ occurs conjointly with state $y_j$ from $\mathcal{L}2$. For instance, $p(x_1,y_1)$ could represent the probability of finding a single-peaked \ion{Mg}{2} spectrum in the same pixel as a blue Doppler shifted \ion{Si}{4} spectrum. 
\begin{figure}[t] 
\includegraphics[width=.5\textwidth]{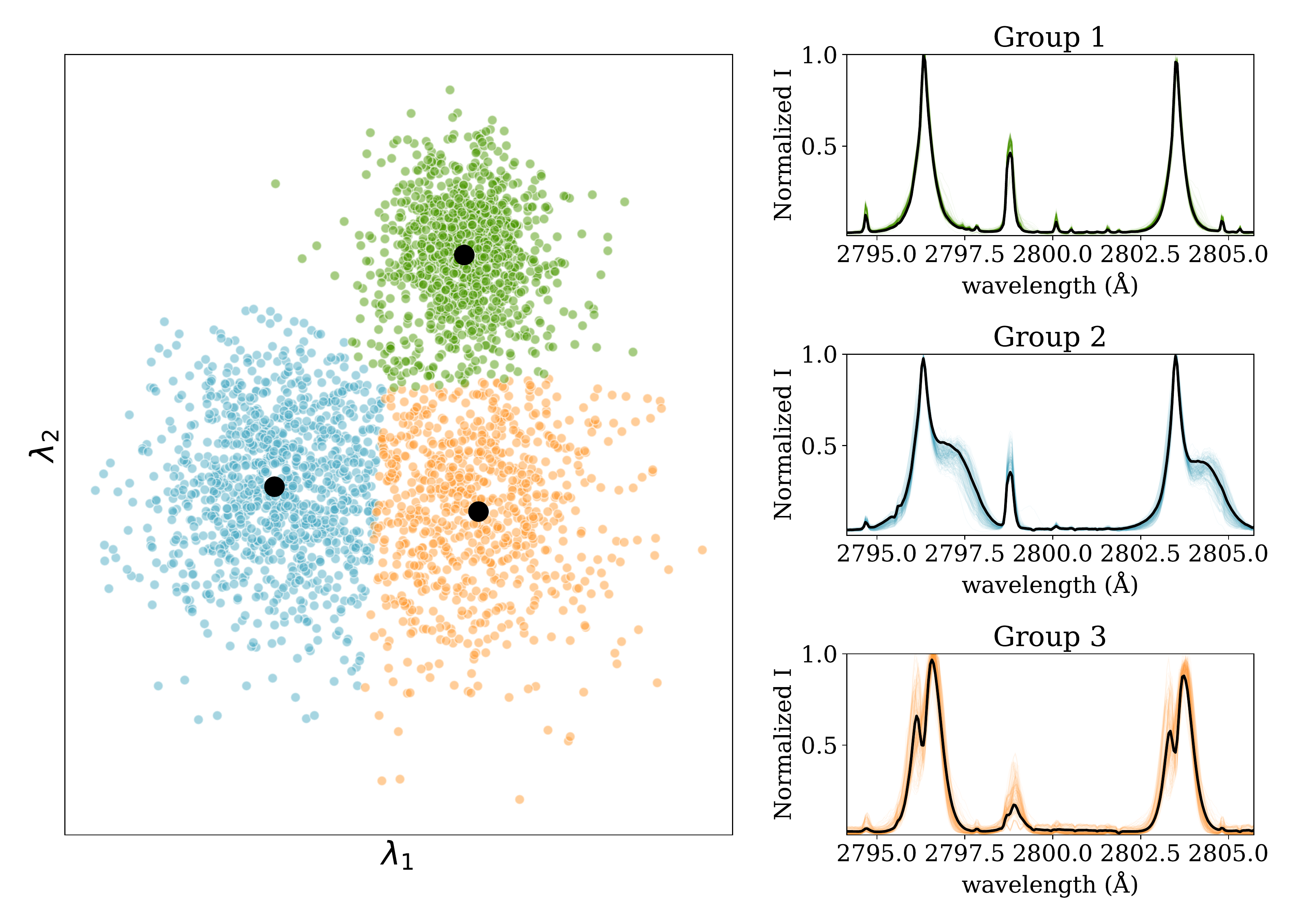} 
\caption{Synthetic representation of a k-means clustering for the \ion{Mg}{2} spectral line. Each point in the left panel represents an entire spectral profile. In reality, the points are in a 240-dimensional space ( onedimension for each wavelength); the plot projects this into 2 dimensions for a visual simplification. After applying the k-means algorithm, the spectra are assigned into 3 groups. The black points represent the centroids or prototypes of each group. The three centroids are plotted on the right in black, with their corresponding group members overplotted in color. Each spectral profile is assigned a label, 1, 2, or 3, depending on which group they belong to, so that k-means imposes a categorical transformation on the data $\mathbb{R}^d\to\mathbb{N}$.}
\label{kmeans}
\end{figure}

The data in its current form makes the construction of this matrix impractical, since the shape of each spectrum can vary continuously, allowing each profile to be distinct. In this case, the number of states for each line equals the number of spectral profiles in the observation, so that we have a trivial one-to-one mapping between $\mathcal{L}1$ and $\mathcal{L}2$, corresponding to a statistically uninformative homogeneous same-colored grid in Figure \ref{prob}. In order to construct $P$, we therefore have to restrict the number of states available for labeling the spectra, and disregard any unimportant details. This is a form of data binning, and is commonly used to estimate the MI of continuous random variables \citep{SCOTT_1979}.

The granularity of the binning is something we have to set \textit{a priori}. The grid that we bin over is defined as $G(k1, k2)$, where $k1$ and $k2$ are the number of unique states for two line-pairs, often referred to as the lines \textit{cardinality} (our example in Figure \ref{prob}  has a grid $G(7,7)$). We have discussed how this binning technique breaks down for asymptotically fine grids, and the same is true for asymptotically coarse grids. If the cardinalities $ki$ are extremely low, then spectra with very different profile shapes may be assigned into the same state, making any physical interpretations untenable. It is therefore desirable to avoid both these limiting cases, however, the limits for which the binning breaks down are uniquely defined for each line and the particulars of the method used to distribute the labels.

After selecting the size of our grid, we have to establish an automatic way of distributing the labels among the spectra, such that spectra assigned the same label (state) also share a similar shape. There are many unsupervised machine-learning techniques that can distribute the set of labels in this intuitive way. We selected the celebrated k-means clustering algorithm of \cite{macqueen1967}. The algorithm initiates points $\mu_j$ called \textit{centroids}, randomly within the data space, and then performs coordinate descent, iterating between an assignment and an update step, with the objective of minimizing the within cluster distances $\mathcal{D}=\sum_{i=1}^{n} \sum_{j=1}^{k} \delta_{c_{i j} \|}\left\|x_{i}-\mu_{j}\right\|^{2}$ between the centroids $\mu_j$ and their assigned spectra $x_i$ \citep[see][for details]{Panos_2018}. Since we normalized out the intensity information in the pre-processing step, the spectra are grouped according to their shape.

In Fig \ref{kmeans}, we have provided a synthetic example of a converged k-means clustering. Each group contains profiles of a similar shape and are represented by the group centroid, which is taken to be the average of all group members. Representing the probability densities with prototype vectors or centroids is commonly referred to as a \textit{vector quantization}.\\

\begin{figure*}[thb]
\centering
    \includegraphics[trim={0cm 2.7cm 1cm 3cm},clip, width=0.33\textwidth]{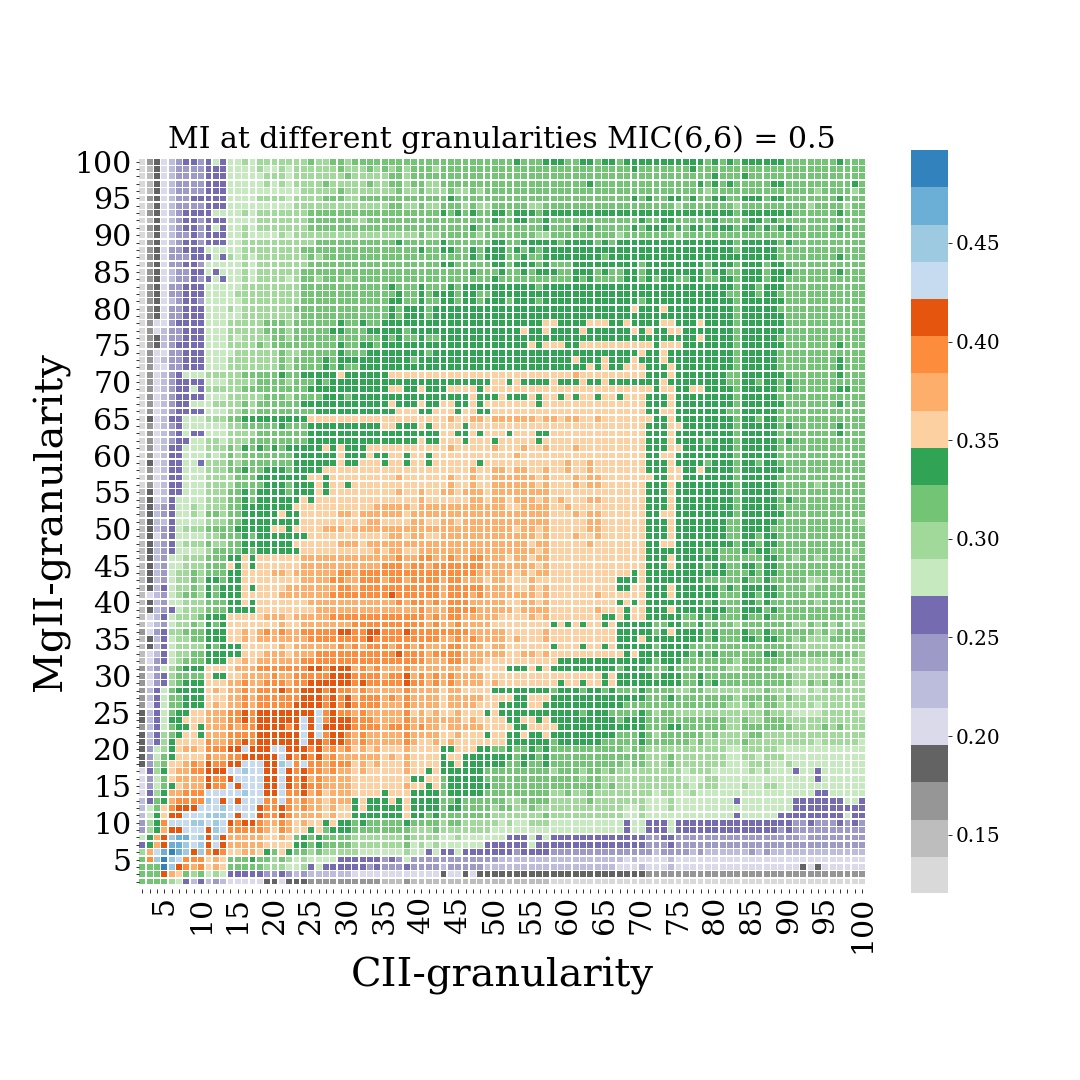}
    \includegraphics[trim={0cm 2.7cm 1cm 3cm},clip, width=0.33\textwidth]{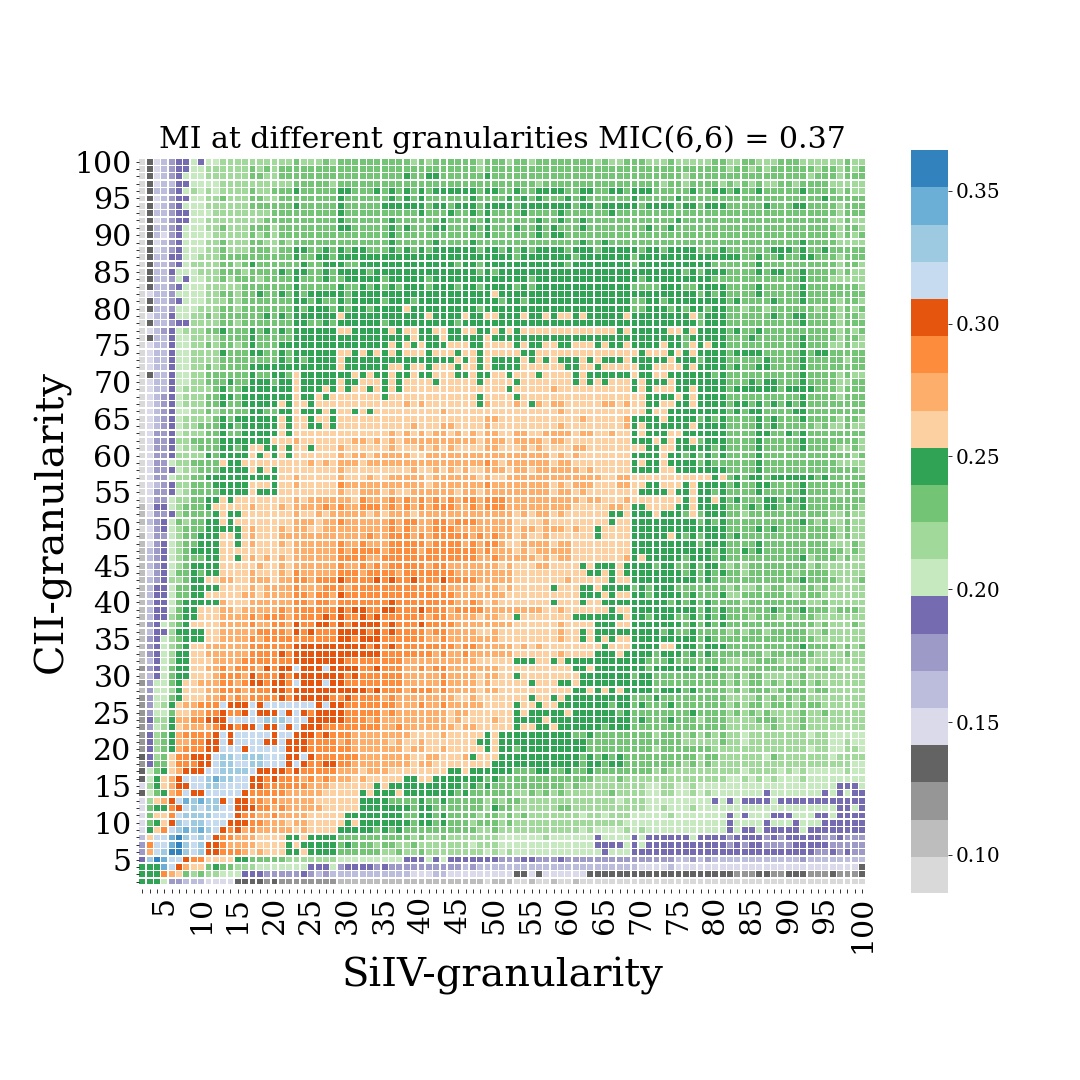} 
    \includegraphics[trim={0cm 2.7cm 1cm 3cm},clip, width=0.33\textwidth]{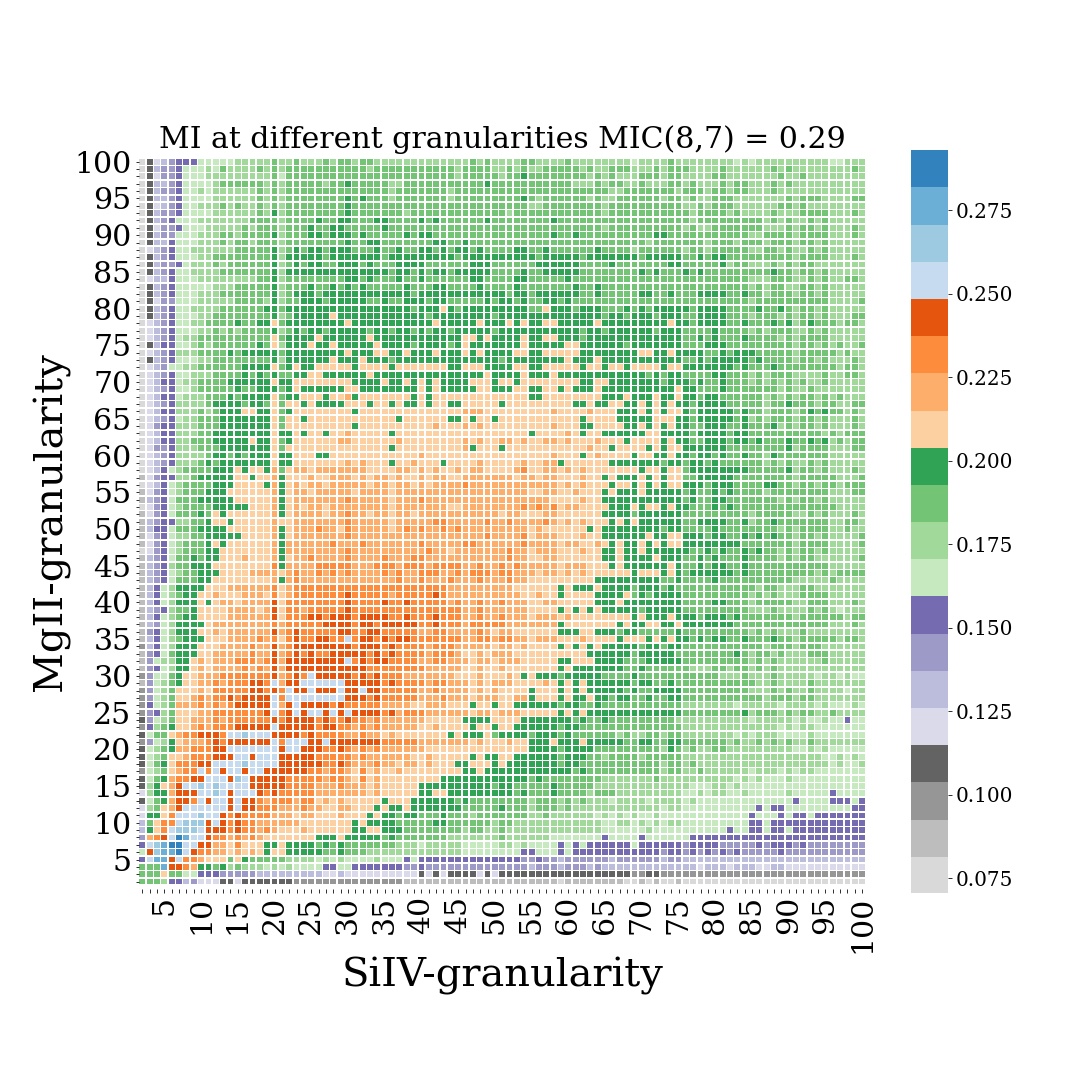}
    \includegraphics[trim={0cm 2.7cm 1cm 3cm},clip, width=0.33\textwidth]{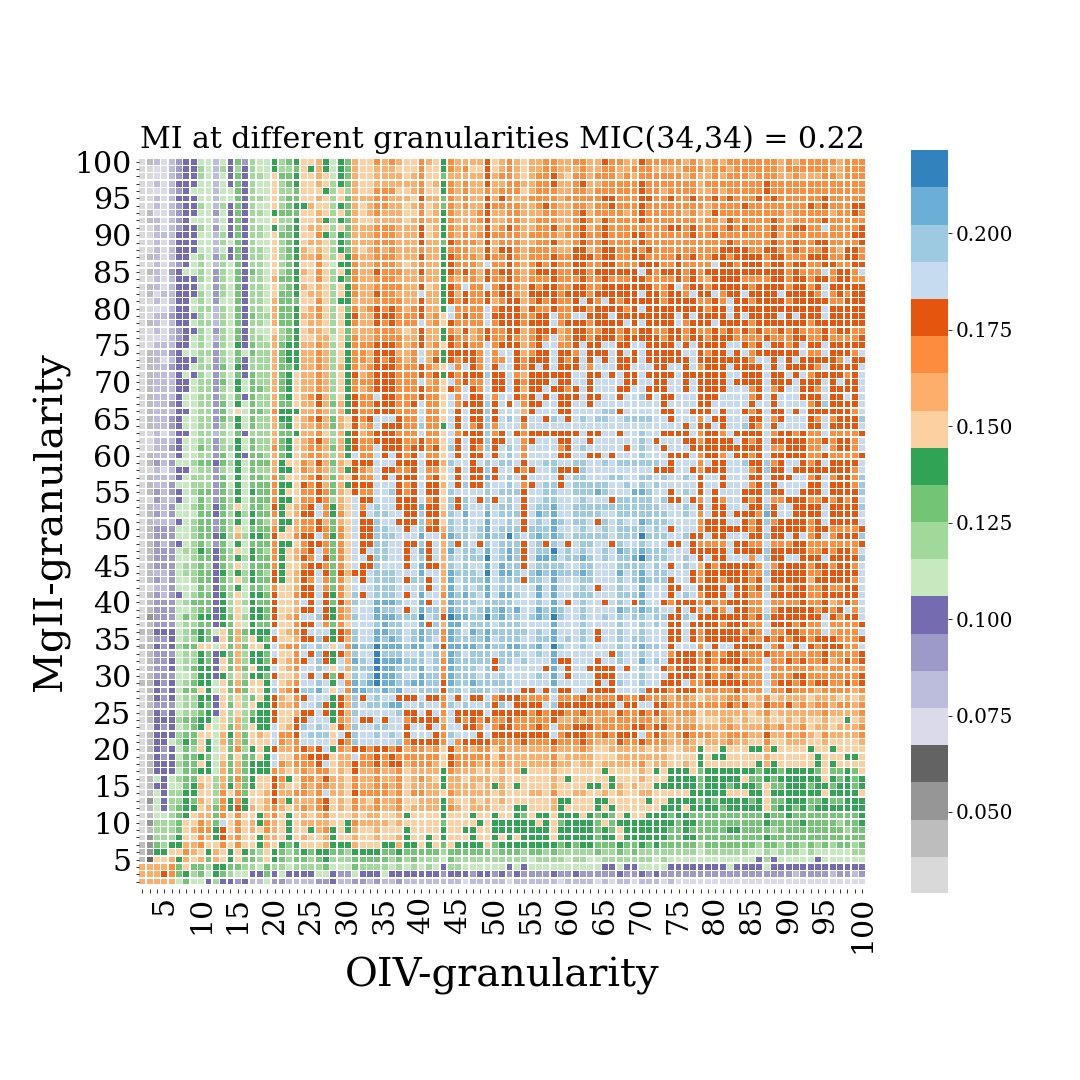} 
    \includegraphics[trim={0cm 2.7cm 1cm 3cm},clip, width=0.33\textwidth]{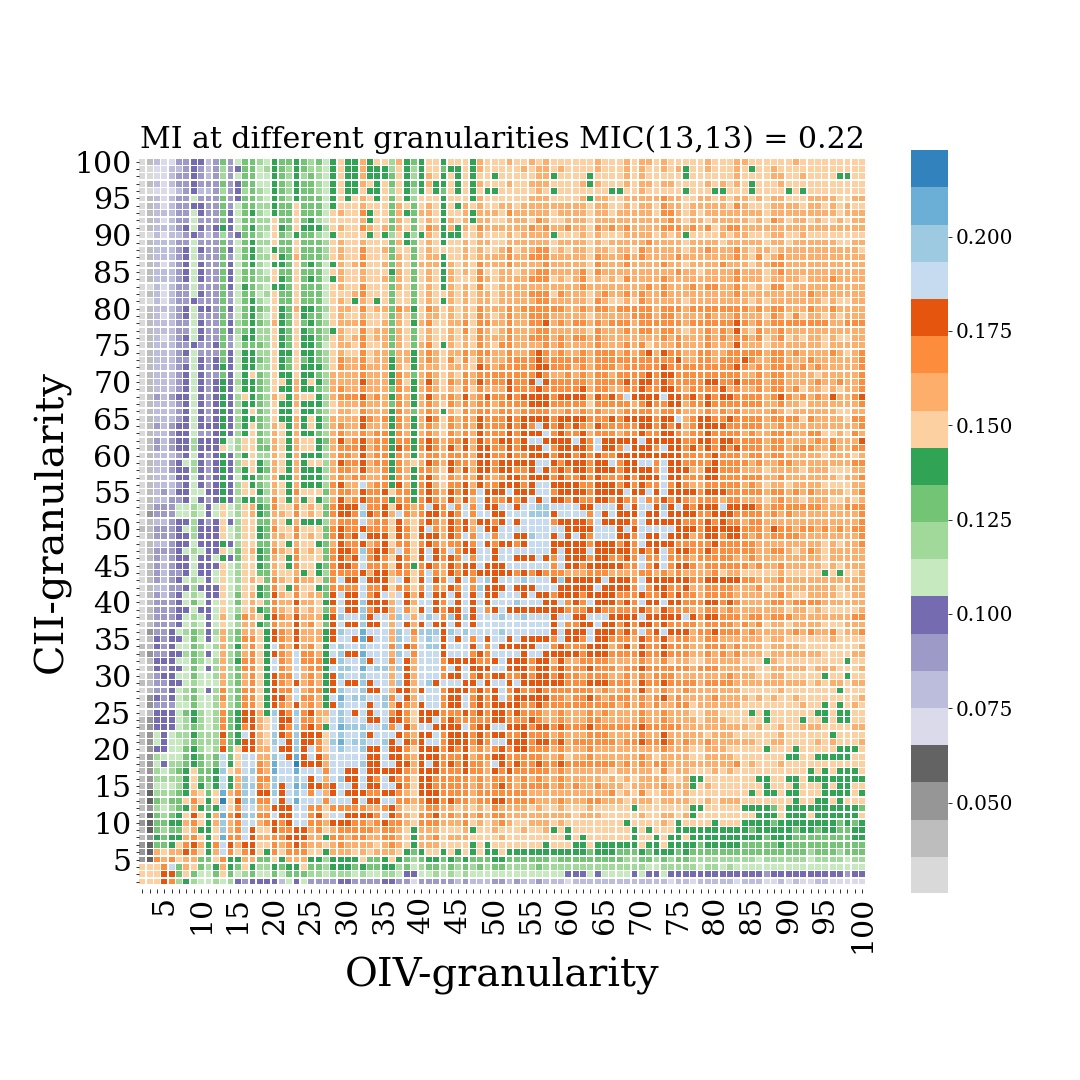} 
    \includegraphics[trim={0cm 2.7cm 1cm 3cm},clip, width=0.33\textwidth]{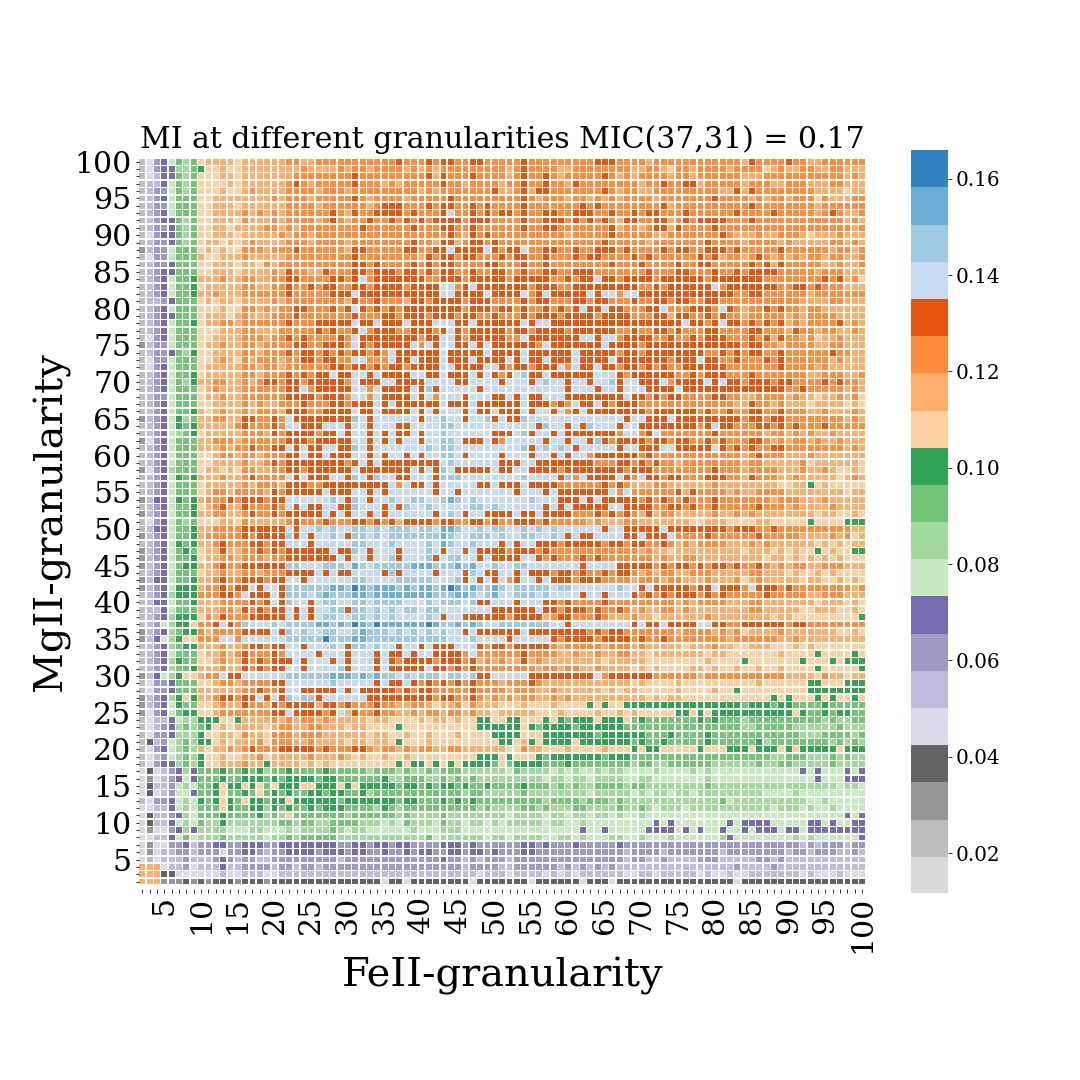} 
    \includegraphics[trim={0cm 2.7cm 1cm 3cm},clip, width=0.33\textwidth]{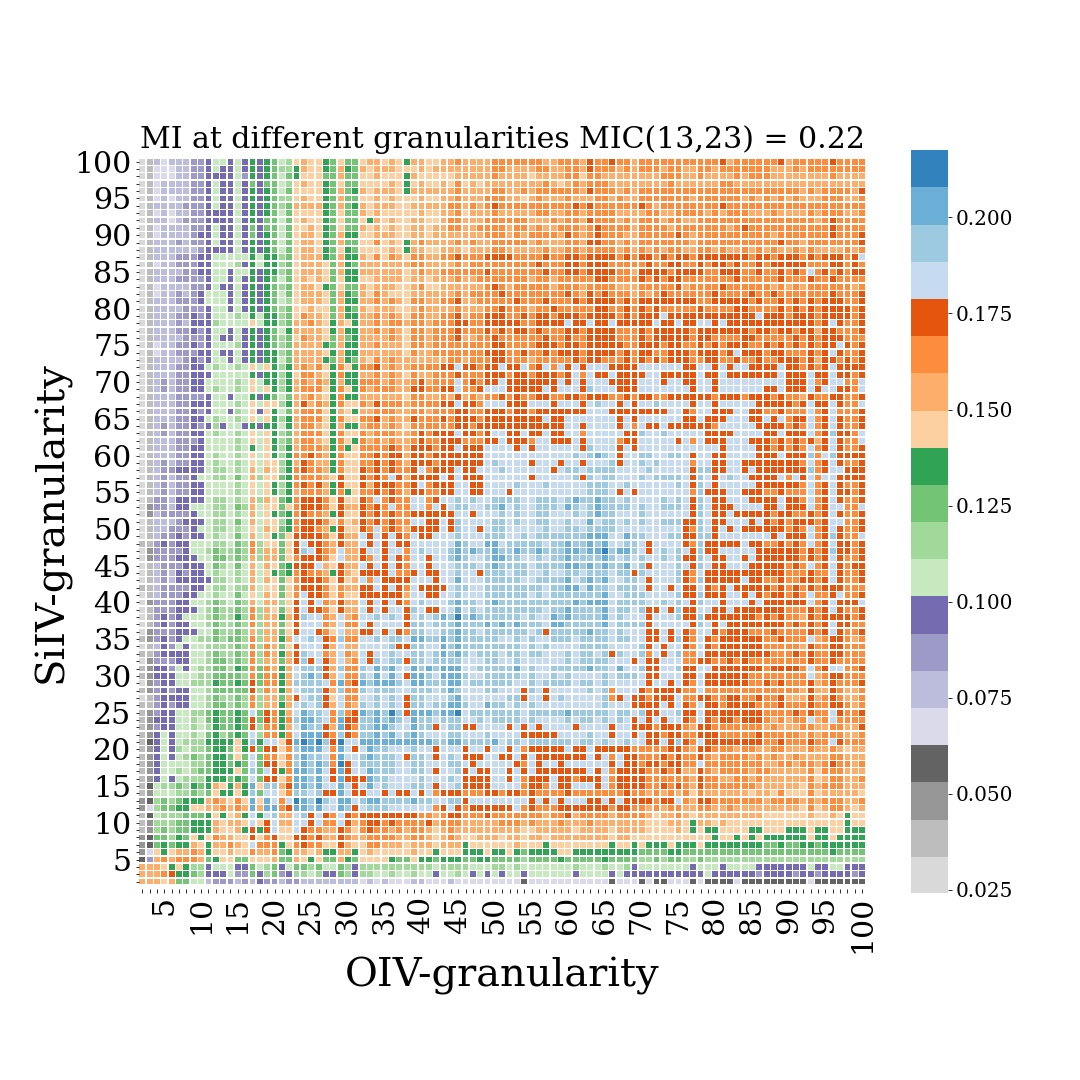} 
    \includegraphics[trim={0cm 2.7cm 1cm 3cm},clip, width=0.33\textwidth]{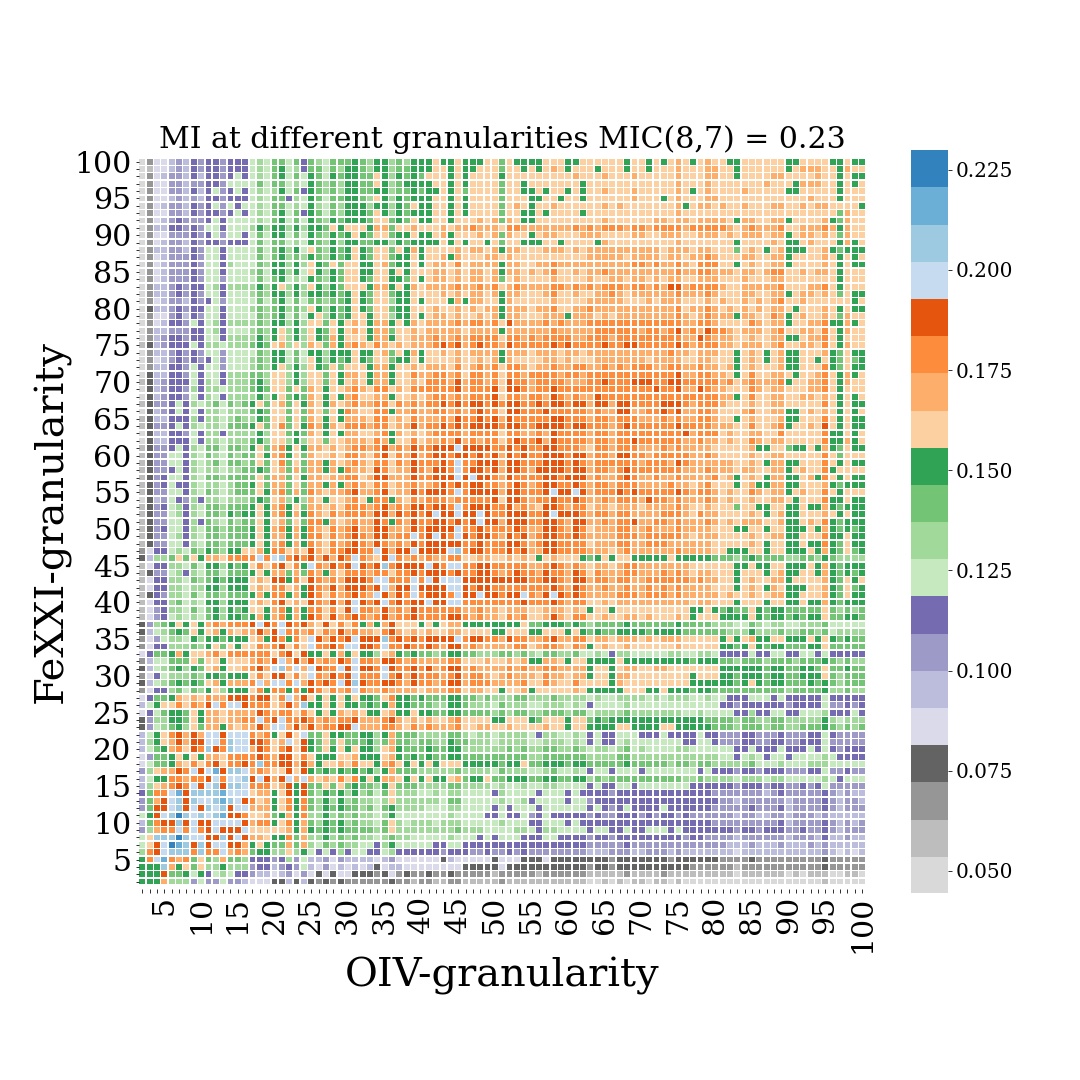}
    \includegraphics[trim={0cm 2.7cm 1cm 3cm},clip, width=0.33\textwidth]{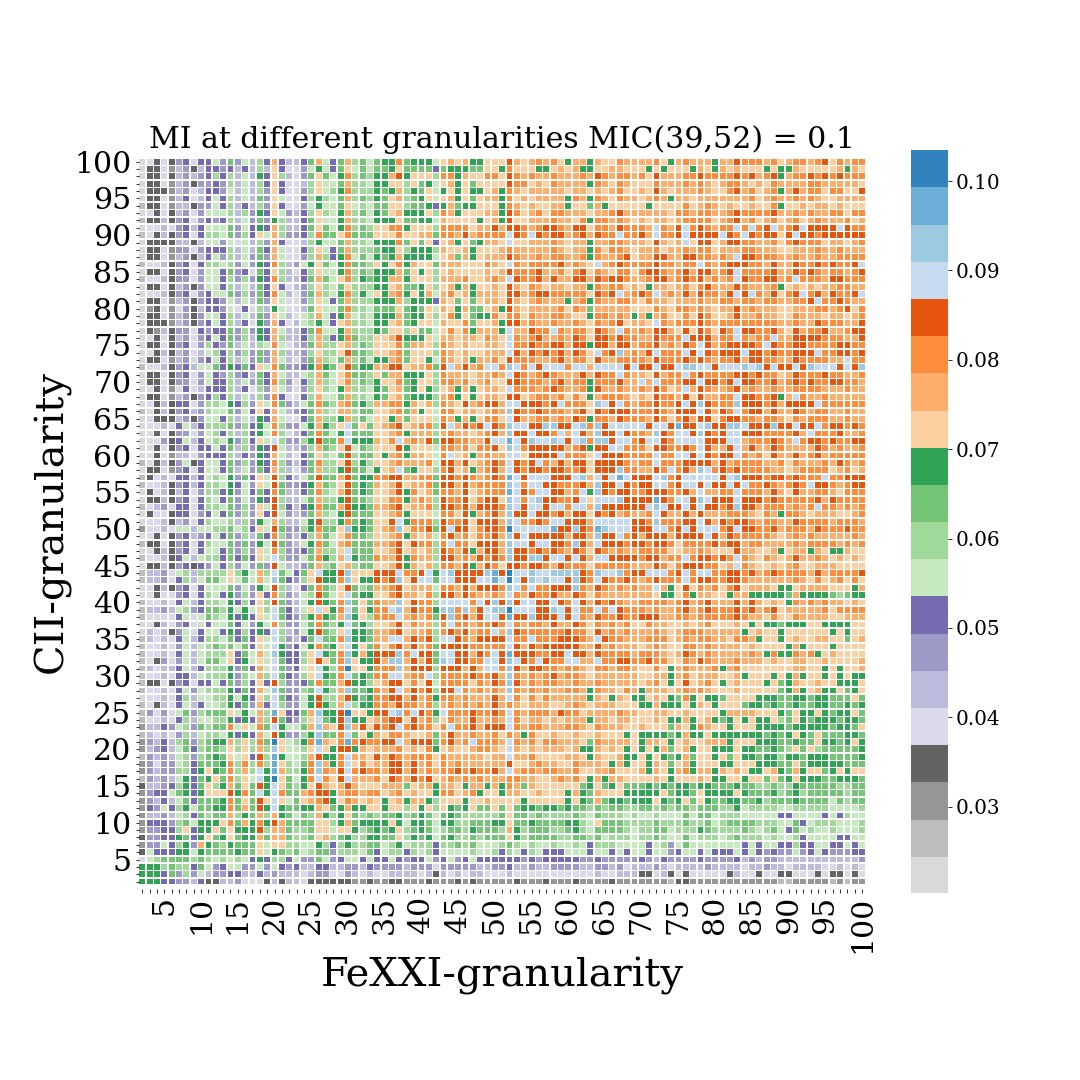}
    \includegraphics[trim={0cm 2.7cm 1cm 3cm},clip, width=0.33\textwidth]{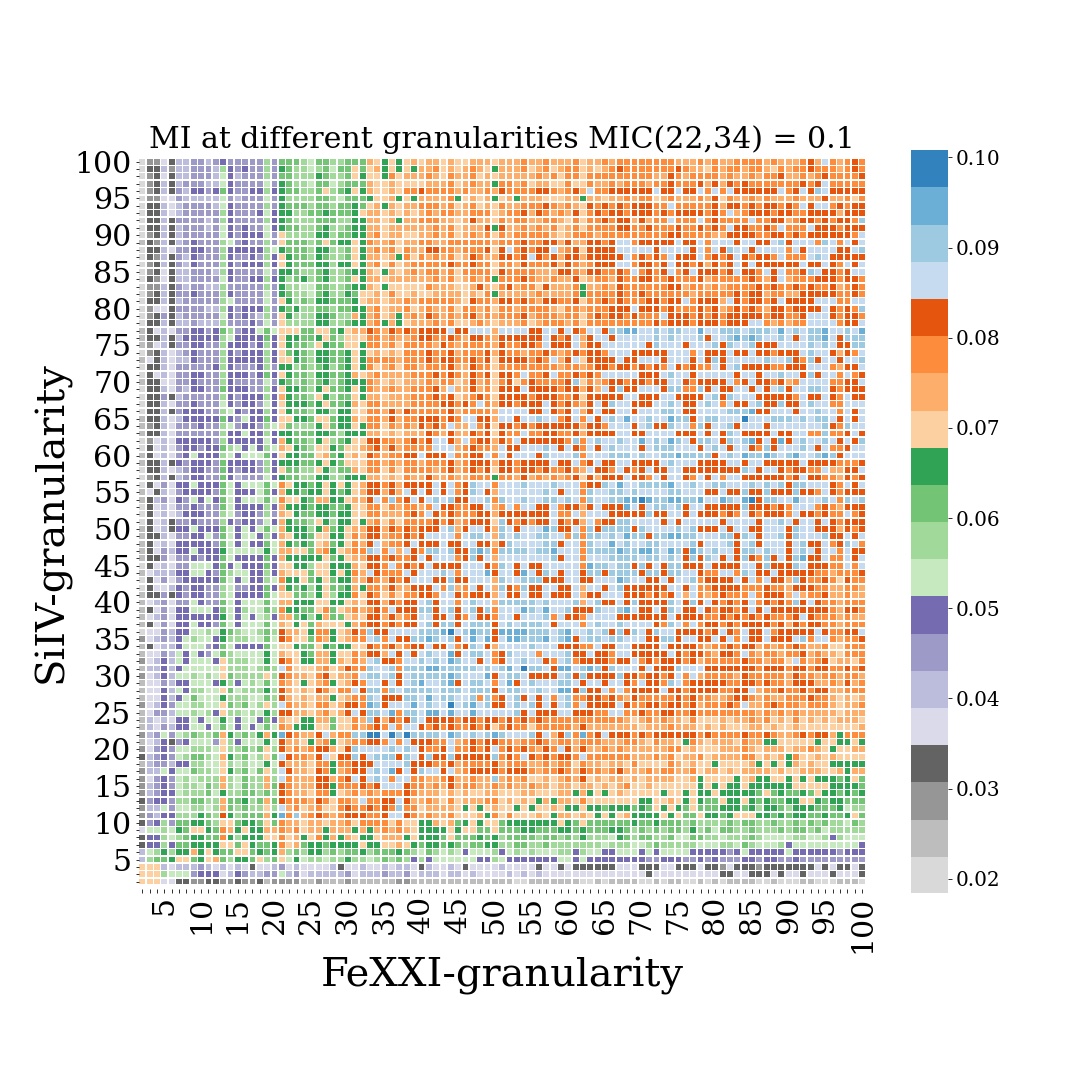} 
    \includegraphics[trim={0cm 2.7cm 1cm 3cm},clip, width=0.33\textwidth]{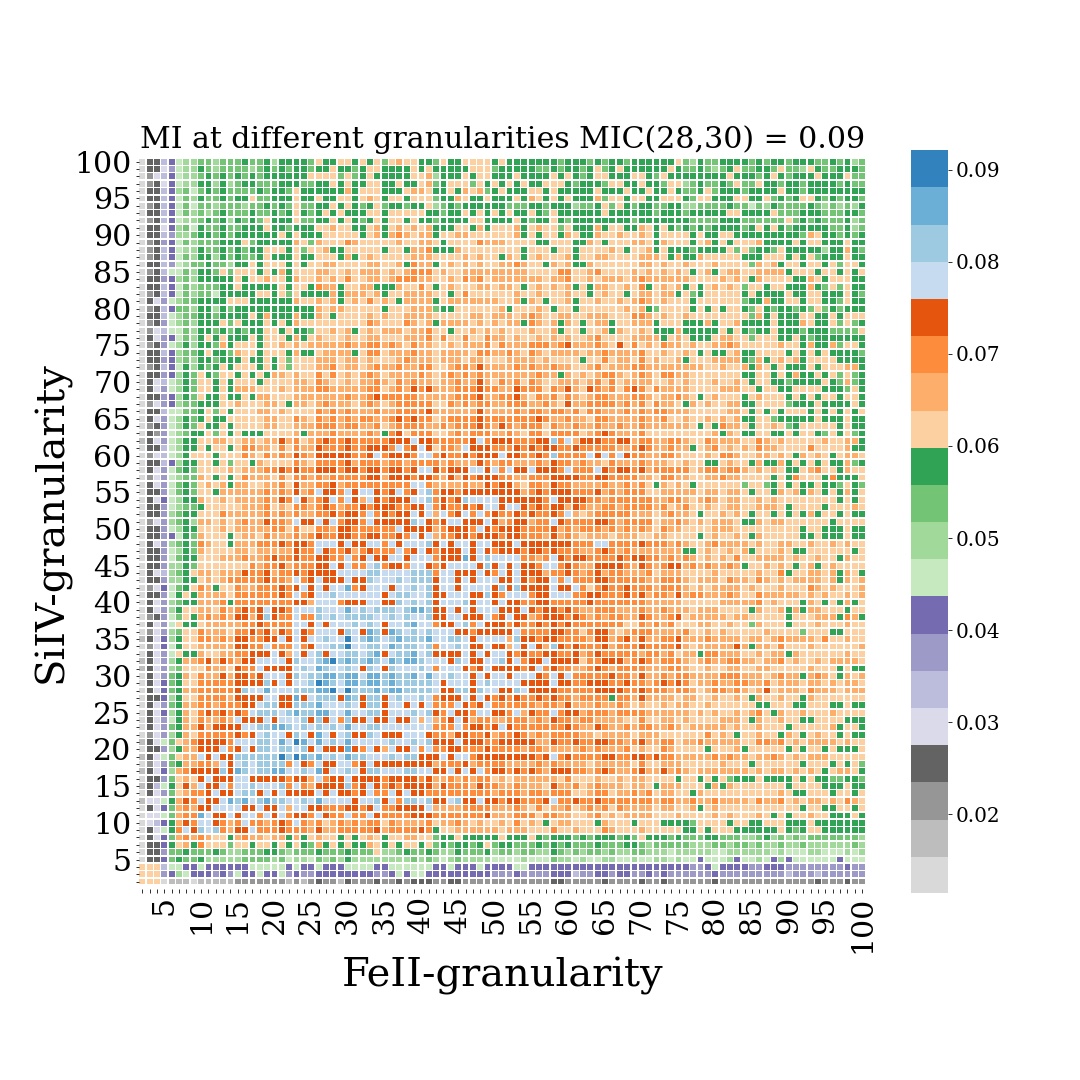}  
    \includegraphics[trim={0cm 2.7cm 1cm 3cm},clip, width=0.33\textwidth]{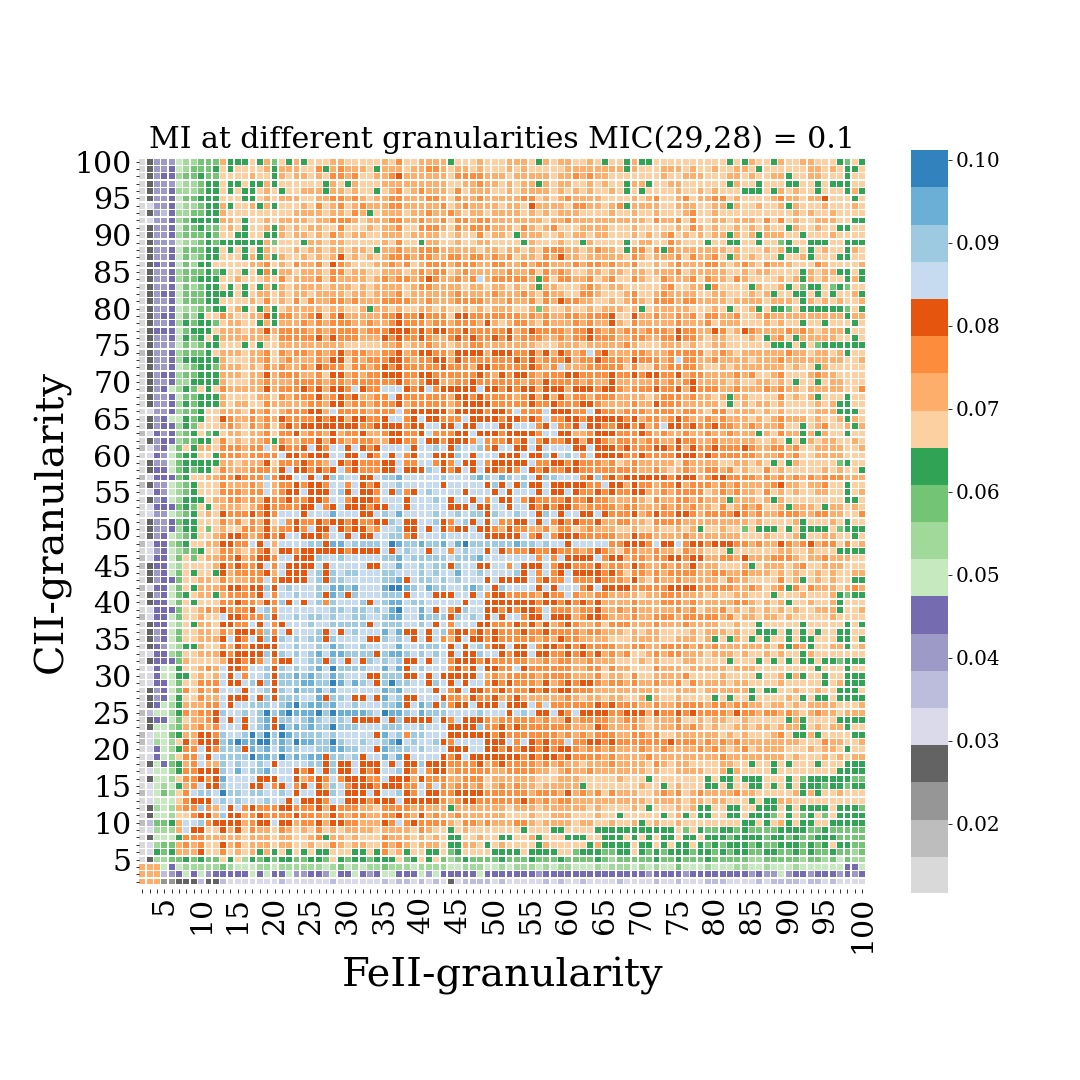}
    \caption{Mutual information $MI_G(X;Y)$ between line-pairs using the categorical approach over the entire flare data set for different grid sizes. Each panel contains MI scores for a particular line-pair, while each pixel represents the MI for a particular choice of granularity. The MIC as well as the optimal choice of cardinalities for each line-pair are indicated in the titles of the subplots. 
\label{MIC}}
 \end{figure*}

 \begin{figure*}[thb]
\centering
    \includegraphics[trim={0cm 2.7cm 1cm 3cm},clip, width=0.33\textwidth]{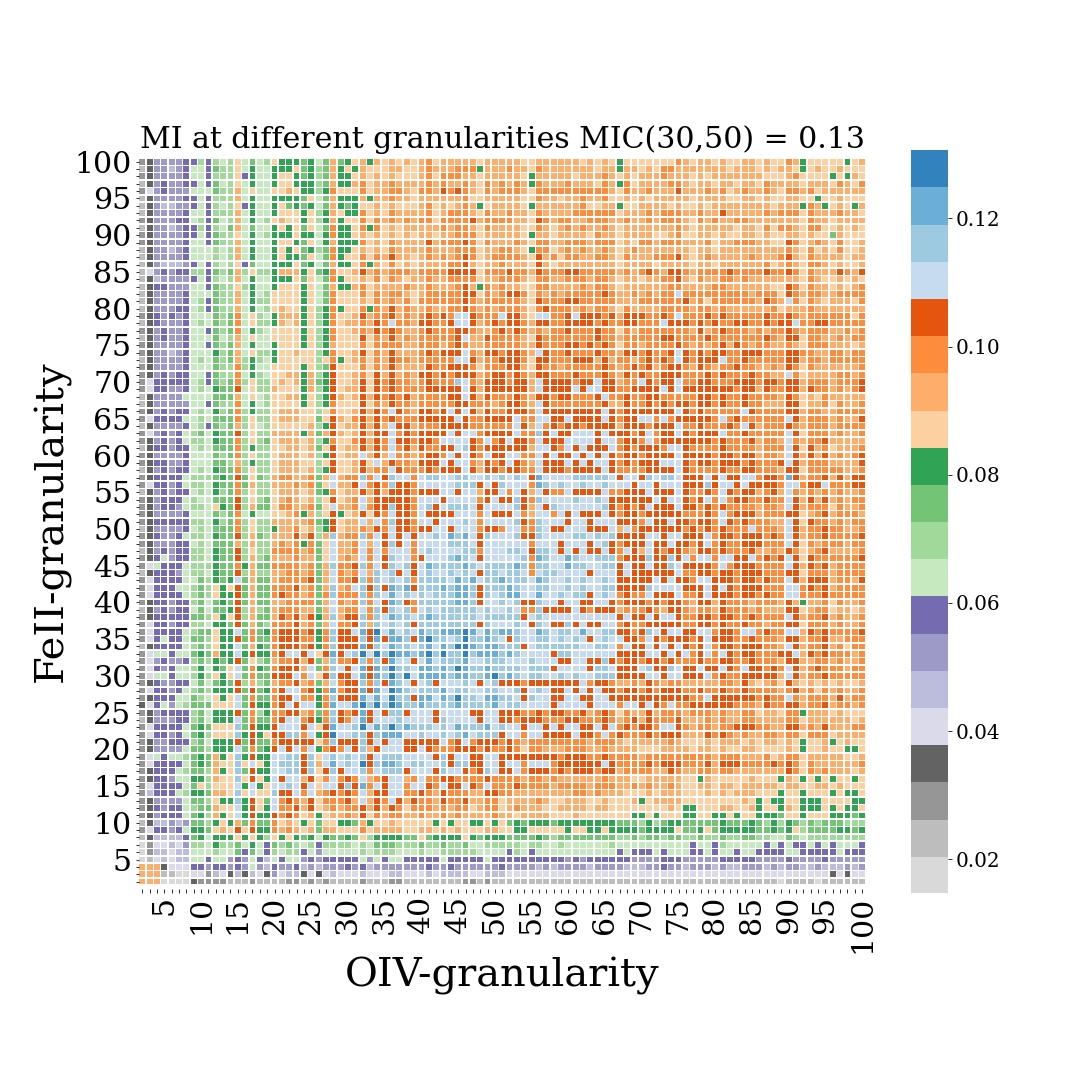}
    \includegraphics[trim={0cm 2.7cm 1cm 3cm},clip, width=0.33\textwidth]{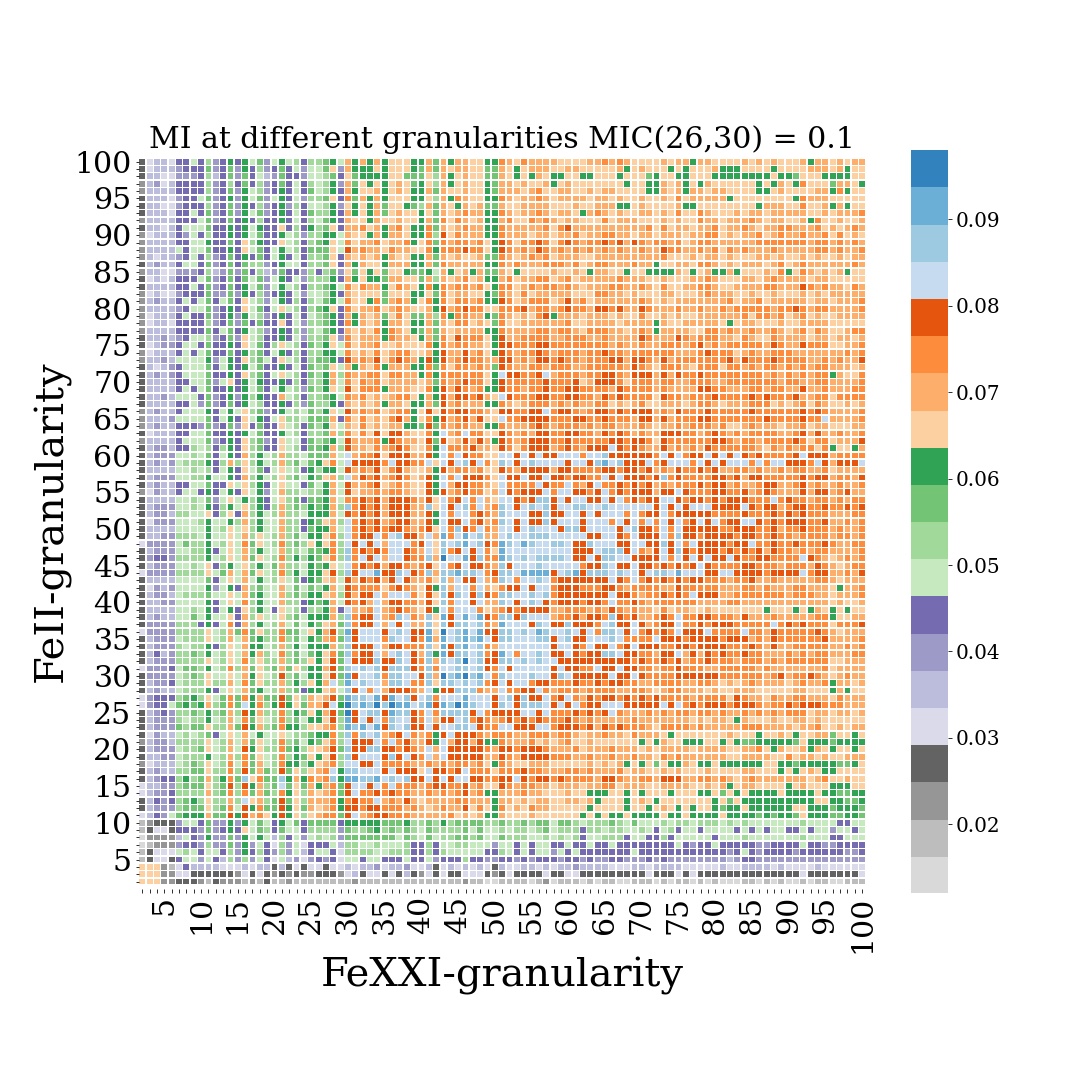} 
    \includegraphics[trim={0cm 2.7cm 1cm 3cm},clip, width=0.33\textwidth]{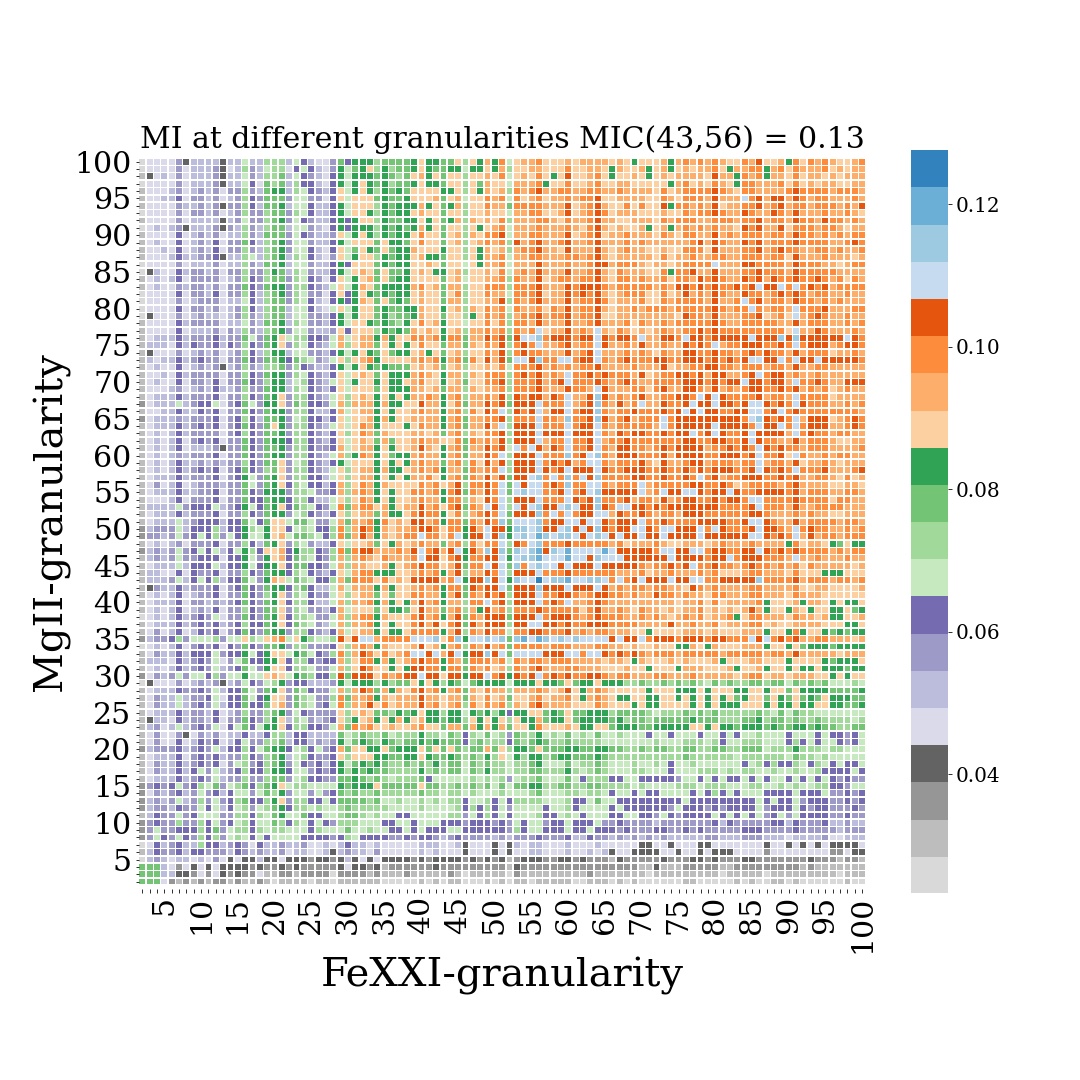}
\caption{Fig.~\ref{MIC} continued.}
\label{MIC2}
\end{figure*}

We used k-means to generate six lists of labels, one list for each spectral line. By comparing the list of labels, say from \ion{Mg}{2} and \ion{Si}{4}, we could generate a true probability distribution $P_{G(k1,k2)}$, calculate both the marginal and conditional probability distributions by summing either over the rows or columns and consequently estimate the MI using Eq.\ref{mutual_information}. Because the MI depends heavily on the particular choice of granularity $G$ (the number of unique labels available for each line), one ordinarily has to repeat the calculation over a large set of different granularities $G\in \mathcal{G}(k1, k2)$, and select the maximum MI. This statistical approach of exhausting all pairs of grid sizes is known as the \textit{maximal information coefficient} ($\mathrm{MIC}$), defined as
\begin{equation}
\mathrm{MIC}\equiv\max\left( \frac{\max _{G \in \mathcal{G}(k1, k2)} MI(X(G) ; Y(G))}{\log \min (k1, k2)} \right),
\end{equation}
where the normalization factor $\log \min (k1, k2)$ penalizes synthetically high MIs associated with fine grid sizes, ensuring comparable results at different cardinalities. The results of this analysis can be seen in Figs.~\ref{MIC} and \ref{MIC2}. Each subplot contains MI scores for a particular line-pair, while each pixel in a subplot represents the MI for a particular choice of grid size. The $\mathrm{MIC}$ as well as the optimal choice of cardinalities for each line-pair is indicated in each subplots title. The advantage of this approach is that one can immediately select a set of pixels with the feature that one is interested in (e.g. single-peaked Mg spectra). A disadvantage is that k-means requires some supervision to verify that the grouping captures all variations that we want to study. We therefore also explored a second method without k-means.

\begin{figure*}[tb] 
\centering
\includegraphics[width=1\textwidth]{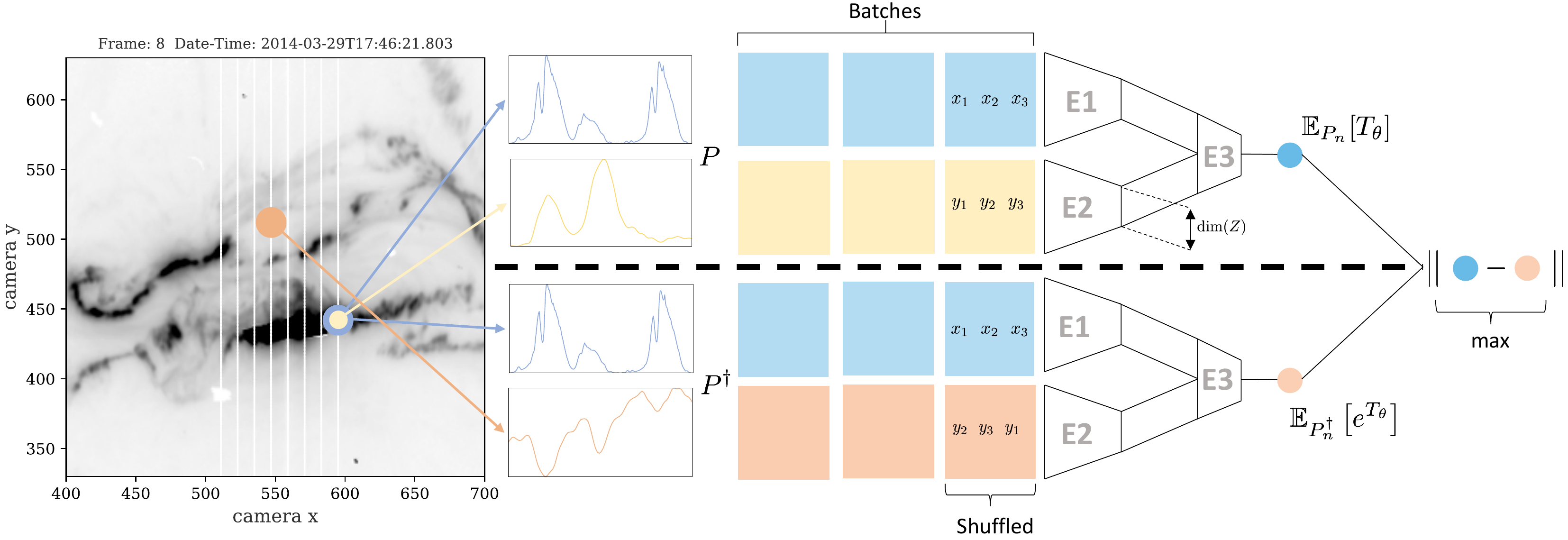} 
\caption{Architecture for the mutual information neural estimator (MINE), used to calculate the correlations between spectral lines directly. The raw spectra from $\mathcal{L}1$ and $\mathcal{L}2$ are initially fed pixel-wise, one batch at a time into the network (blue and yellow), where a batch refers to a small subsample of the data used to update the networks parameters. The spectra are then compressed by encoders $E1$ and $E2$ into a latent representation, where they are subsequently concatenated and compressed again by $E3$ into a real number represented by the blue circle. The same data with a permutation on $\mathcal{L}2$ (blue and orange), is then processed in an identical manner to produce another real number represented by the orange circle. Since the upper half of the network samples from the true distribution, and the lower half from the independent distribution, the real number outputs can be substituted into the objective function given by Eq.\ref{ML_mine}, which in turn is used to train and eventually estimate the MI. The degree to which the network can separate the blue and orange circle is directly proportional to the degree of correlation between the spectral lines. Note that the parameters between the upper and lower half of the network are shared, i.e., encoders $E1,2,$ and $3$ in the top half are the same encoders used in the bottom half.}
\label{mine_net}
\end{figure*}

\subsection{Approach 2: Numerical}
  \label{num_sec}
  
We now develop an approach that allows us to estimate the MI directly from the raw continuous spectral data, without reverting to a categorical transformation, and without enlisting an additional clustering algorithm to distribute labels. This provides us with an objective result that does not depend artificially on any particular choice of hyperparameter. Because of the technicality of this method, we will first provide a brief summary of the intuition behind the idea, after which we will substantiate the method mathematically.\\

 \subsection{Idea behind the approach}
Suppose that we receive pairs of spectral shapes from two different lines with similar formation heights. Furthermore, suppose that both $\mathcal{L}1$ and $\mathcal{L}2$ are sensitive to line-of-sight velocity flows, which are encoded as simple Doppler shifts from their core rest wavelengths. A human could tell if the two lines were correlated or not by inspecting individual pixels. If the magnitude and direction of the Doppler shifts for both lines are always proportional within the same pixel, then there is a good chance that the lines are correlated. The connections between spectral lines do not have to be as obvious as the ones discussed here; for instance, the \ion{Mg}{2} subordinate line emission may be connected to downflows within the \ion{Si}{4} line. Indeed there may be a large number of obscure connections that a human observer would not consider analyzing. To address this issue, we allow a neural network to discover a set of optimal features in much the same way that we allowed the VAE in section \ref{VAE_sec}. For the case of the VAE, our objective function was to minimize the difference between real and reconstructed quiet Sun spectra. After training the VAE, we were left with a very useful bi-product called a model, which contained a rich internal description of all interdependence's between the quiet Sun $\lambda$ points. We now replace the objective of minimizing the difference between real and reconstructed spectra with the new objective of maximizing the difference between pairs of spectra from line $\mathcal{L}1$ and $\mathcal{L}2$ sourced from the same, and different pixels. It is important to note that whenever we take $\mathcal{L}1$ and $\mathcal{L}2$ from the same pixels (pixel-wise), we are sampling from the true distribution $P$, as apposed to sampling from the independent distribution $P^\dagger$ (across-pixels). In this way, we construct two data sets, one containing pixel-wise concatenated spectra of $\mathcal{L}1$ and $\mathcal{L}2$, and the other containing concatenations across-pixels. These two streams of data are then fed into a network which tries to distinguish between them. The logic is simple: If we assume that pairs of spectra from different pixels are completely uncorrelated, then the only way for the network to distinguishing the two streams of data, would be to learn all of the dependencies that exist between spectra from the same pixels. The degree to which the network can separate the data sets is then directly proportional to the amount of correlation that exists between the lines, and provides us with an indirect way of monitoring the degree to which $P$ deviates from $P^\dagger$. The biproduct in this case is a model that contains the statistical distributions of intensity interdependence's between spectral lines.\\ 

This procedure is depicted in Figure \ref{mine_net}. The network is separated by a horizontal dashed line. The upper half of the network is fed spectra pixel-wise from line $\mathcal{L}1$ (blue) and $\mathcal{L}2$ (yellow), while the lower half of the network is fed the same data but with a random permutation of the spectra from $\mathcal{L}2$ (orange). The upper half is therefore sampling data from $P$ while the lower half samples data from $P^\dagger$. The network's objective is to try and distinguish between these two streams of data. Focusing on the upper half of the network: each pair of spectra are passed to an encoder (represented by $E1$ and $E2$), which compresses the profiles down into a latent space of dimension $|z|$. The features from both lines are subsequently concatenated and passed to a final encoder $E3$, that maps the $2|z|$ features into a single real number represented by the blue circle. The same procedure maps the data from the lower half of the network into a real number represented by the orange circle. The encoders here are similar to the encoder used in the VAE of Figure \ref{VAE_arch}, in that they find an optimal basis (set of features) that best minimize the objective function. These features are referred to as latent variables because they are not preprogramed but emerge organically as solutions during the training process. The network's objective is now to separate the blue and orange circle as much as possible. If there is no difference between pairs of pixel-wise spectra and shuffled spectra, then the network will not be able to distinguish the two streams of data, resulting in zero separation between the two circles. On the other hand, if the two streams of data are distinctly different from one another, then the network will be able to leverage this difference and separate the blue and orange circles. In this way, the degree to which the circles are separated can be used as a measure of correlation. \\ 

Notice again the parallel between this method and the previous method in section \ref{cat_sec}. In both cases, the true $P$ and independent $P^\dagger$ distributions are estimated, and the dependence is subsequently calculated as the difference between them. It can be shown that if we require the network to maximize the difference between the two circles, then the two real numbers associated with these circles can be used to estimate the MI with arbitrarily high accuracy. We will now substantiate this claim and demonstrate how the network in Figure \ref{mine_net} flows naturally from a numerically compatible reformulation of the $\mathrm{KL}$-divergence.

\subsection{Mathematical description}
\cite{Donsker_1983} showed that there exists a function $T$, such that Eq.~\ref{KL} can be written in a way that engages expectation values from both probability distributions, giving us the so-called dual representation
\begin{equation}
D_\mathrm{KL}(p\| q)=\sup _{T: \Omega \rightarrow \mathbb{R}} \Big\{ \mathbb{E}_p[T]-\log \left(\mathbb{E}_q\left[e^{T}\right]\right) \Big\},
\label{Donsker}
\end{equation}
where "sup" stands for the supremum, meaning that the equality only holds for the particular function $T^*$, which maximizes the expression. As usual, the MI appears as the special case where $p\to P$ and $q\to P^\dagger$. Simply stated, the dual representation says that we can measure the difference between the true $P$ and independent $P^\dagger$ distributions indirectly by measuring the different effects they have on a third distribution. Concretely, we form a weighted sum of some random distribution $T$ with respect to both the grid values of the true and independent distributions. The difference between the two summed versions of $T$ encodes something about the difference between $P$ and $P^\dagger$. It is easy to see that some functions have a higher capacity for reflecting this difference than others. For instance, a distribution with a support that coincides only with the equal value grid points of both $P$ and $P^\dagger$ will not be capable of highlight the discrepancies that may exist. It follows that any suboptimal function $T$ will only ever result in an underestimation of the actual dependence, i.e., $D_\mathrm{KL}(P\| P^\dagger)\geq D_\mathrm{KL}(P\| P^\dagger ; T)$.\\

The dual formulation has therefore afforded us a numerical foothold into a variational maximization problem that did not exist before. We can initiate an arbitrary distribution $T$, and update it in such a way as to maximize $D_\mathrm{KL}(P\| P^\dagger; T)$, while always being assured that the MI can never be overestimated. Of course we cannot explore an infinite number of distributions; however, the expressive power of NNs allows us to sample the most relevant fraction of these functions \citep{Universal}. Therefore, instead of taking the supremum over all possible functions, we consider the subset of functions $\mathcal{F}=\{T_\theta\ | \theta\in\Theta\}$, parameterized by an NN, and rewrite Eq.~\ref{Donsker} for the case of MI as:
\begin{equation}
MI(X;Y)_{n} \geq \sup_{\theta\in\Theta} \Big\{ \mathbb{E}_{P_n} \left[T_{\theta}\right] - \log\left(\mathbb{E}_{P^\dagger_n }\left[e^{T_{\theta}}\right]\right) \Big\}.
\label{ML_mine}
\end{equation}
The possible functions are explored in a controlled and directed manner, using backpropagation to update the network's parameters after each batch of spectral samples. Since the network is slowly trained over subsamples of data, each $n$th batch provides a different estimate of the true underlying distributions; hence, the MI is parameterized by the batch number $n$. The fact that NNs are universal function approximators ensures that the value after the algorithm converges will approximate the actual MI with arbitrary accuracy.\\

In summary, we can sample from the true $P$ and independent $P^\dagger$ distributions by collecting spectra either pixel-wise or across pixels. The MI is given by the difference between these two distributions. This difference can be obtained indirectly by monitoring the individual effects of $P$ and $P^\dagger$ on a third distribution $T$. Some $T$ distributions are more suitable than others. If we allow a NN to be the function which generates $T$, then we can numerically optimise $T$ until it has the capacity to reflect most, if not all of the differences between $P$ and $P^\dagger$. The statement in Eq.~\ref{ML_mine}, combined with the fact that NNs are universal function approximators, ensures that the result after convergence will be a tight lower bound on the actual MI. What we have discussed here is precisely the network shown in Figure \ref{mine_net}. This particular type of network is called a mutual information neural estimator (MINE), developed first by \cite{MINE_2018}.

Recent investigations by \cite{mcallester2019formal} have contended the assumption of a tight lower bound for the particular case of a large MI and small sample size, showing that methods that maximizes a variational lower bound, such as MINE, cannot obtain a bound larger than $\mathcal{O}(\log n)$, where $n$ is the number of samples. Therefore, there is the potential for MI to be grossly underestimated in such a circumstance.\\

Some of the most exciting features of statistical networks like MINE are the low computational demands, short training times and sparse coding schemes given the current advanced state of high-level deep learning libraries. For our particular task, the network consisted of  32 lines of python code written within the PyTorch library \citep{PyTorch}. This code includes both the network architecture and training loop. The training time required to estimate the MI between a single line-pair under flaring conditions including both orange and yellow pixels, required the processing of $2332884\times 2$ spectral profiles, which running on a GeForce RTX 2080 Ti TURBO 11G for 20 epochs took 2 minutes. The computational demands can be further decrease if one uses \textit{stochastic gradient descent}, where instead of incrementally updating the networks parameters after each epoch, we update the network parameters using estimations of the true gradient from small samples of data. We found that the network could be trained on a CPU with a similar time complexity using a batch size and number of iterations of 1000. 

\begin{figure}[t] 
\centering
\includegraphics[width=.5\textwidth]{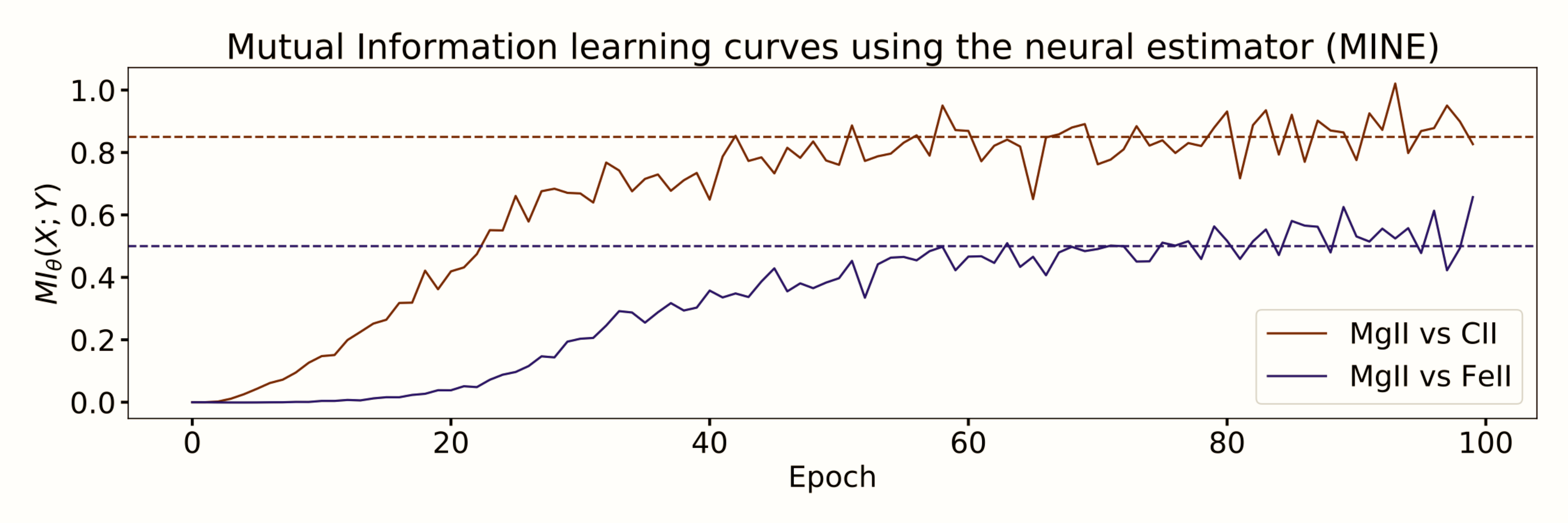} 
\includegraphics[width=.5\textwidth]{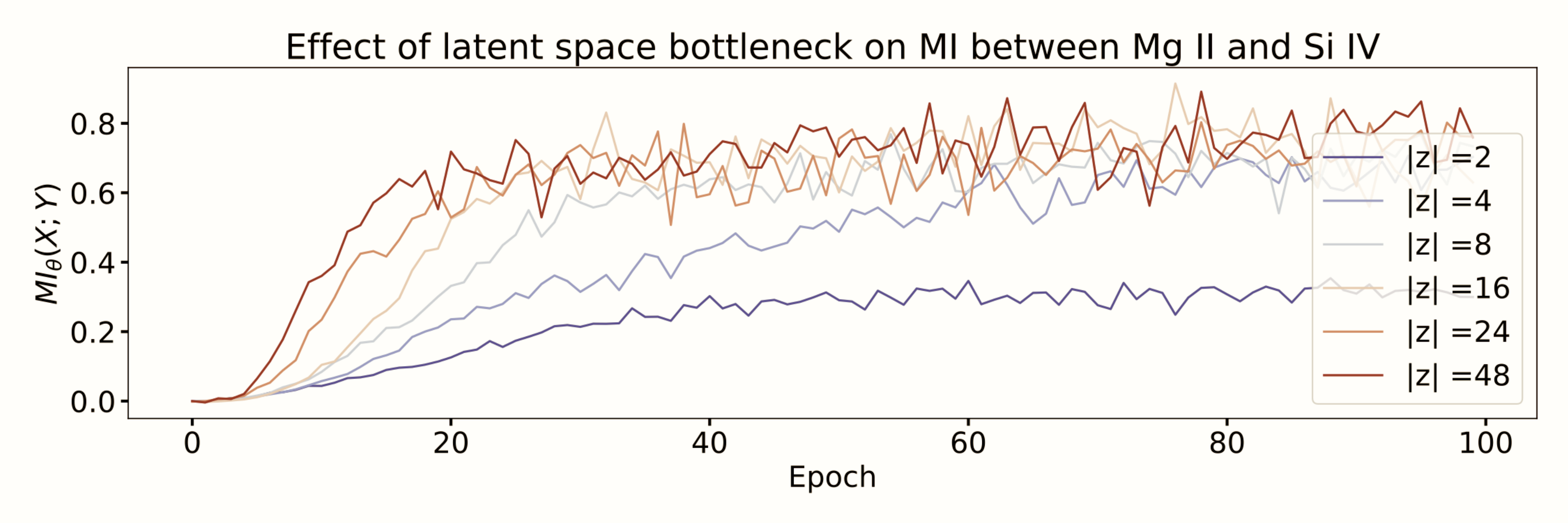} 
\caption{Learning curves for the estimation of line-pair mutual information via the application of MINE. Upper panel: MI curves for two different line-pairs. The horizontal dashed lines indicate the values of MI at convergence. The mutual information in this case is not normalized, and the line-pair spectral data were retrieved from a single flare observation. Results indicate that \ion{Mg}{2} is more strongly correlated with \ion{C}{2} than with \ion{Fe}{2}. Lower panel: effect of latent space bottleneck on the convergence of mutual information. For latent dimensions greater than 2, the network has enough expressibility to capture the salient features of each line and saturate. This implies that, unlike the categorical method, the MI from MINE does not depend on the choice of any hyperparameter. Furthermore, any additional intricacies introduced into the architecture of the network, such as convolutions, dropouts, or pooling, only affect the rate of convergence.}
\label{latent_space}
\end{figure}

\begin{figure}[t]
\centering
    \includegraphics[trim={0cm 0cm 0cm 0cm},clip, width=0.23\textwidth]{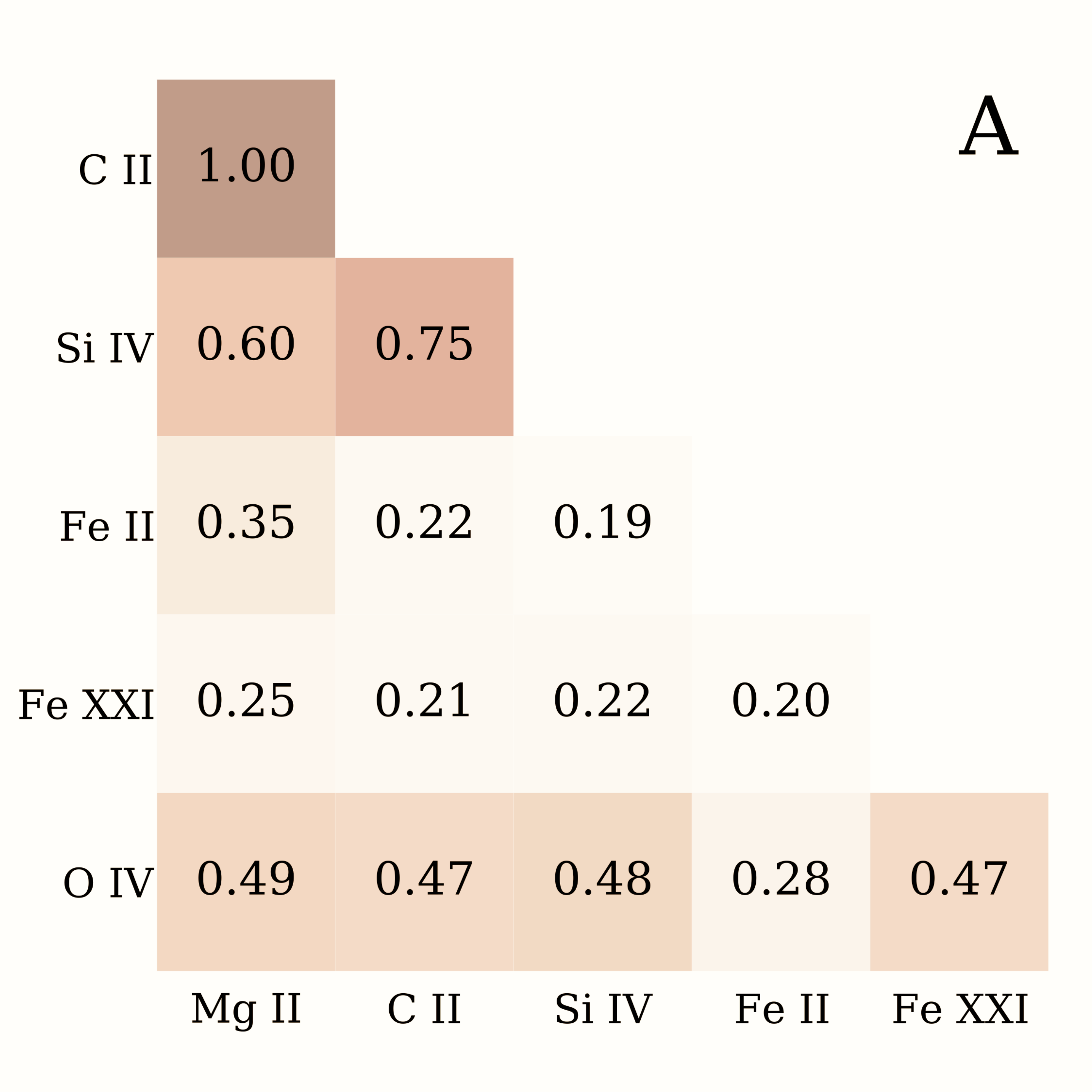}
    \includegraphics[trim={0cm 0cm 0cm 0cm},clip, width=0.23\textwidth]{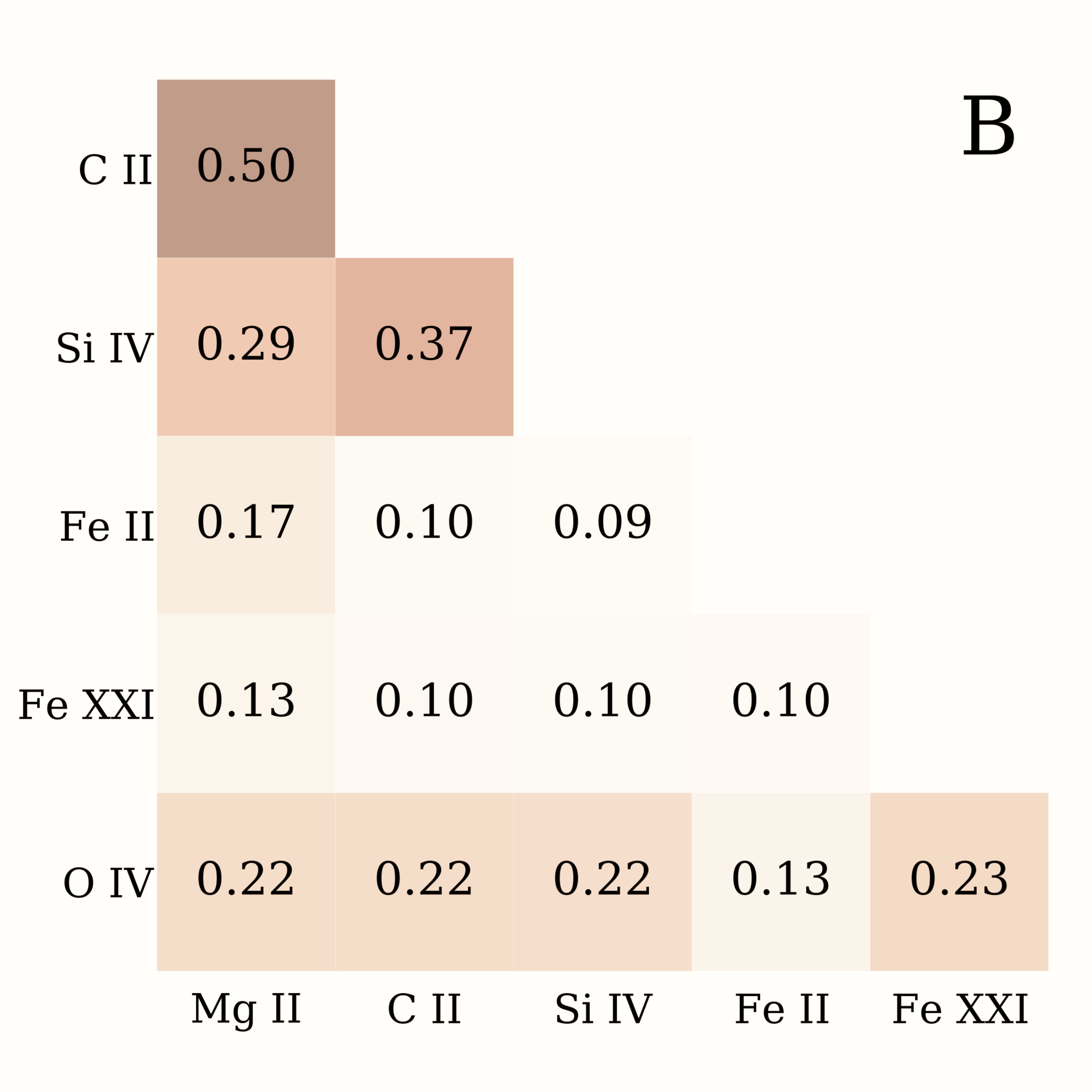}
    \includegraphics[trim={0cm 0cm 0cm 0cm},clip, width=0.23\textwidth]{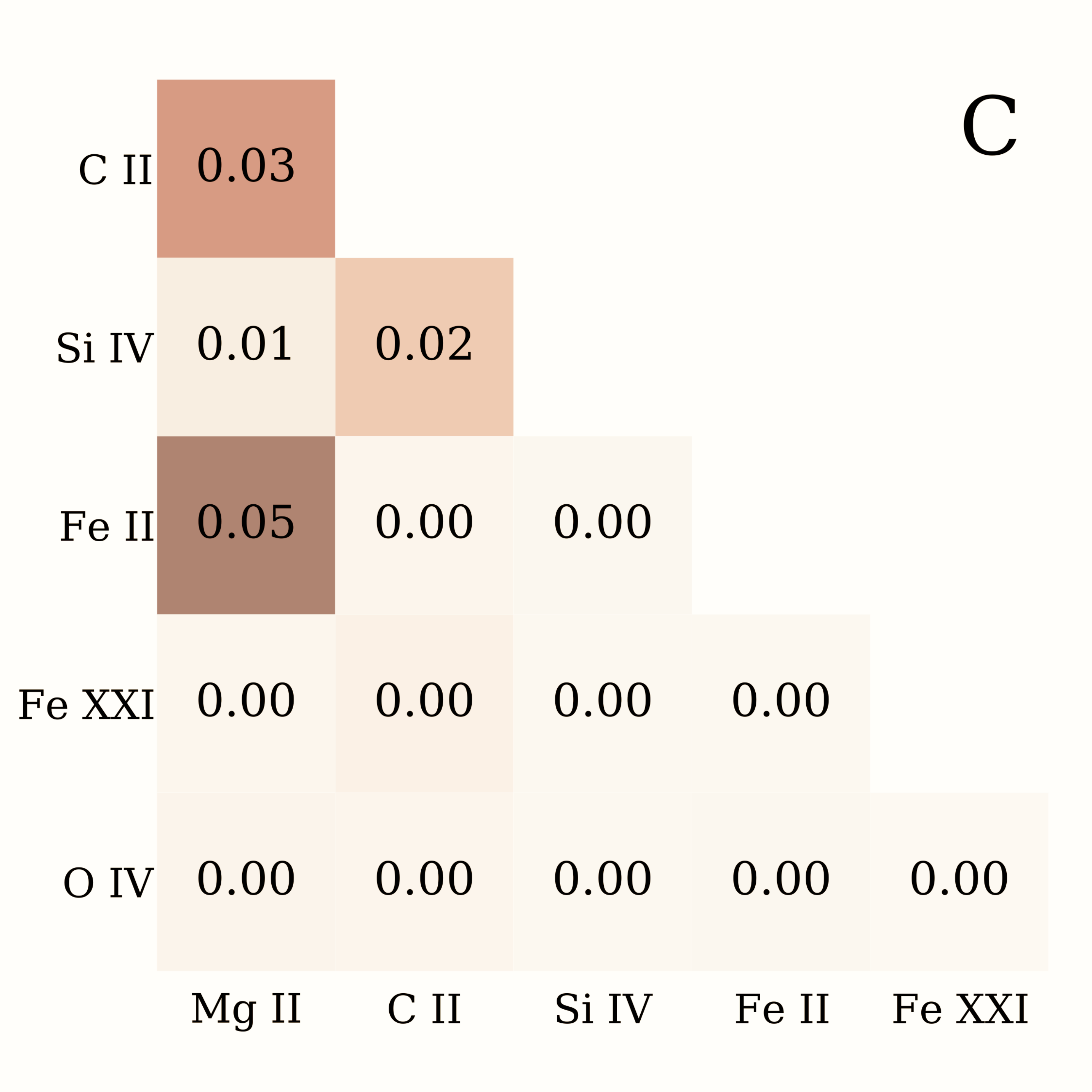} 
     \includegraphics[trim={0cm 0cm 0cm 0cm},clip, width=0.23\textwidth]{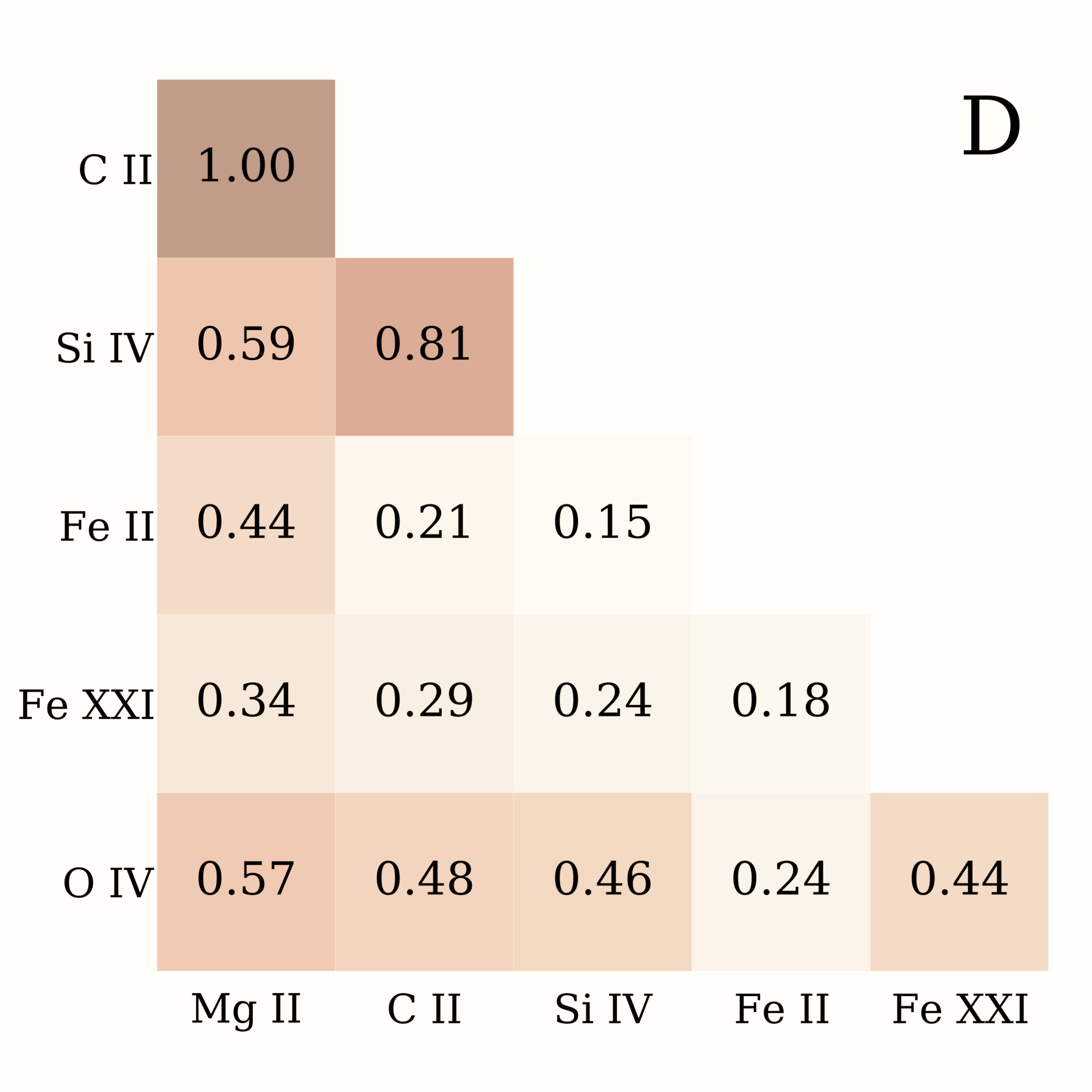} 
\caption{Panel A: relative mutual information, calculated using the categorical method and integrating over each grid in Fig \ref{MIC}-\ref{MIC2}. Each MI is normalizing by the maximum line-pair MI. These results indicate that the two chromospheric lines \ion{Mg}{2} and \ion{C}{2} have the strongest correlation, followed by the hot \ion{Si}{4} line. Panel B: normalized mutual information for the entire flare data set at optimal MIC cardinality settings. \ion{Mg}{2} and \ion{C}{2} show substantial couplings at granularities (6,6), while the iron lines produce spectral profiles that are largely independent. \ion{O}{4} shows signs of weak line coupling. Panel C: normalized mutual information between line-pairs for the entire quiet Sun data set at optimal MIC cardinalities. Results indicate that line coupling turns off during quiet Sun conditions, with little to no dependence between profile shapes. Panel D: relative MI between flaring line-pairs using the MINE network. Results are consistent with the categorical results in panel A.}
\label{Res}
\end{figure}

\begin{figure*}[t]
\centering
    \includegraphics[trim={0cm 0cm 0cm 0cm},clip, width=1\textwidth]{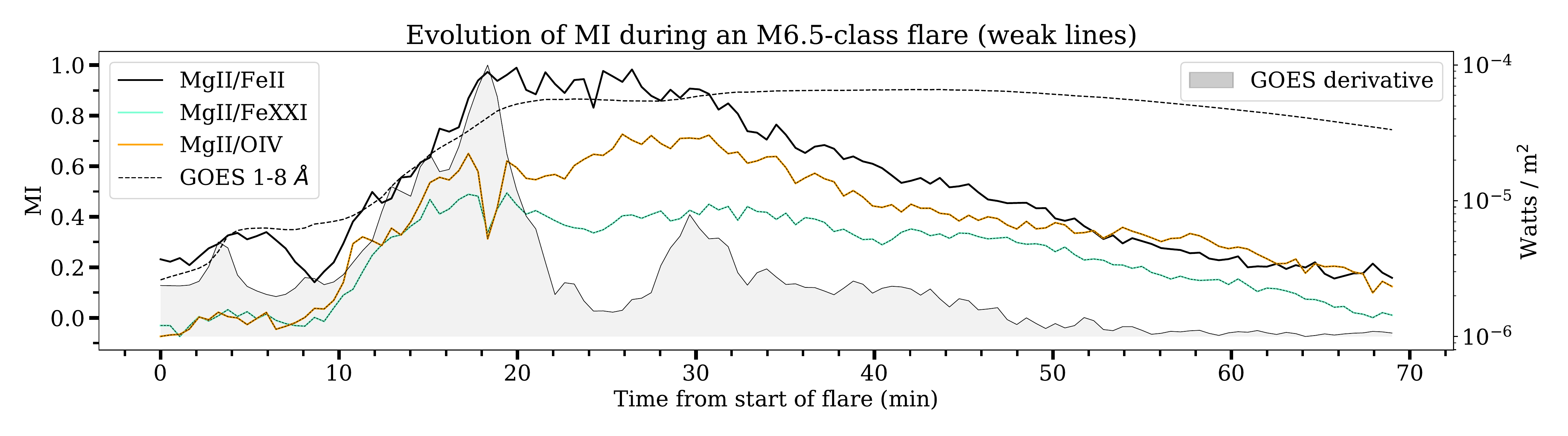}
\caption{Temporal evolution of the correlation between \ion{Mg}{2} and the three weaker lines: \ion{Fe}{2}, \ion{Fe}{21} and \ion{O}{4} during a large M6.5-class flare. The MI was calculated using the MINE network from spectra collected over each of the $129$, 16-step rasters. The flare analyzed here is flare 13 in Table \ref{obs}. Along with the MI, we have included the GOES soft X-ray curve and derivative. The weak lines follow a close relationship with the X-ray signature of GOES, and couple strongly to magnesium during the flare. This is in conflict with the comparatively low scores indicated in Figure \ref{Res}. Additionally, the flaring region occupied the entire FOV of IRIS. We postulate that the weaker lines only become coupled over the most energetic regions of the solar flare, possibly over the flare ribbons where the signal-to-noise ratio increases. In contrast, for the MI between magnesium and the strong lines, \ion{C}{2} and \ion{Si}{4} remain highly coupled throughout the entire flare (not shown here). Note that because the MI was calculated using the MINE network, the MI is not required to be bound between 0 and 1. The fact that the \ion{Fe}{2} correlation peaks at 1, is therefore  a coincidence, and should not be interpreted as a perfect correlation.}
\label{EV_MI}
\end{figure*}

\section{Results}
\label{Results_section}
We tested the network for $(\text{\ion{Mg}{2}} | \text{\ion{C}{2}})$ and $(\text{\ion{Mg}{2}} | \text{\ion{Fe}{2}})$ on spectra from flare 17 of Table \ref{obs}. The training of MINE for both line-pairs can be seen in the upper panel of Figure \ref{latent_space}, with latent dimension $|z|=10$. The network initially produces suboptimal $T_\theta$ distributions which result in underestimations of the MI, however, with each successive pass over the data set (epoch), the MINE network slowly learns to distinguish between pairs of spectra sampled across-pixel and pixel-wise. The true MI represents the point at which the network converges and stops improving. This point is indicated by the blue and red horizontal dashed lines for \ion{Fe}{2} and \ion{C}{2} respectively. The result indicates that for this single flare observation, the \ion{Mg}{2} line is more correlated with \ion{C}{2} than with \ion{Fe}{2}, with a MI of 0.82 versus a MI of 0.45. Note that only the relative differences are important.

To evaluate the effect of the network's architecture on the results, specifically the dimensionality of the latent space, we performed calculations for a single line-pair (\ion{Mg}{2}|\ion{Si}{4}) at a variety of different $|z|$ settings. The results are encapsulated in the lower panel of Fig \ref{latent_space}. We see that the network is robust against changes to the architecture, with latent latent dimensions $|z|>2$  always converging to the same final MI. We incorporated additional complexities into the network such as dropout and pooling layers. More complex variants of the MINE network appear to only enhance the rate at which the network converges. In conclusion, the MINE network appears to be extremely robust, and given enough time, will always produce an accurate estimate of the true MI.\\

The main results for both the categorical and numeric approaches can be found in Figure \ref{Res}. Each panel displays the estimated MI for different data sets and approaches. Panel A contains the relative MI results from the categorical approach applied to the entire flare data set (including both orange and yellow pixels from the VAE mask). Each index represents the relative MI between the coinciding spectral lines indicated by the rows and columns. The results were calculated by integrating over each grid in Figure \ref{latent_space}, and normalizing by the maximum line-pair MI, which in this case is the \ion{Mg}{2}/\ion{C}{2} line-pair. Panel B represents the categorical results over the entire flare data set, but instead of integrating over each grid, the MIC values for optimal granularity were used, and the normalization between line-pairs was suppressed. Panel C is identical to B, except applied over the entire quiet Sun data set. It shows that line correlations are very weak in the quiet sun. Finally, panel D encapsulates the relative MI scores of the MINE network applied to the entire flare data set (orange and yellow pixels). Although we included the actual numerical values of the MI for the categorical method in panel B, we do not include the MINE equivalent here, since the network does not easily lend itself to estimates of the specific line entropy's, which would ordinarily lead to easily interpretable mutual information's bound between 0 and 1. Even though the absolute values of MI between methods 1 and 2 are not comparable, the order of correlated lines is largely preserved between both methods, which can be seen as similar shades in Panels A and D. 

\subsection{Are weakly coupled lines really weak?}
There is a clear distinction in panel B of Figure \ref{Res}, between lines that couple strongly and lines that couple weakly. In the prior case we have: \ion{Mg}{2}, \ion{C}{2} and \ion{Si}{4}, with the largest correlation for the  (\ion{Mg}{2}|\ion{C}{2}) line-pair, reaching a normalized MI of 0.5. The weakly coupled lines include: \ion{Fe}{2}, \ion{Fe}{21} and \ion{O}{4}. There are two components that could contribute to weak couplings: 1) The above lines are simply not correlated with any other line. This could be the case for the \ion{Fe}{2} line which forms at a height (the photosphere) that is significantly different form the remaining lines. 2) The lines are weak and have a low signal-to-noise ratio. This may be the case for the \ion{O}{4} and \ion{Fe}{21} lines.

In both of the above mentioned cases, it may be possible that weakly coupling lines become strongly coupled at a particular time or over a particular region. For instance, the line couplings could be enhanced during the impulsive phase when IRIS is observing directly over the flare ribbons. This could be in part due to the expected strong bulk velocity flows, which may knit the upper and lower atmosphere together, or from a general enhancement of the quality of the signal-to-noise ratios. To explore these possibilities, we performed a case study on flare \#13 in Table \ref{obs}. This flare was selected because the double flare ribbons unfold during the course of the observation to encompass the entire 16-step raster FOV. Therefore, if the weak lines are only weak because of the particular region we typically observe (regions off of the ribbon), we should in this case expect them to become enhanced. We calculated the MI between \ion{Mg}{2} and the three above-mentioned weak lines for each of the observations raster sweeps. This generated a temporal evolution, as seen in Figure \ref{EV_MI}, of the correlations over 120 rasters.

The results indicate that the MI of the "weak" lines experience a drastic fluctuation during the passage of the flare, and that this fluctuation is directly proportional to the GOES derivative. In fact, all three of the weak lines follow the GOES derivative closely, confirming our suspicion that correlations appear to be enhanced during the impulsive phase over regions receiving a high amount of energy flux. \\

\begin{figure*}[t]
\centering
    \includegraphics[trim={0cm 0cm 0cm 0cm},clip, width=0.32\textwidth]{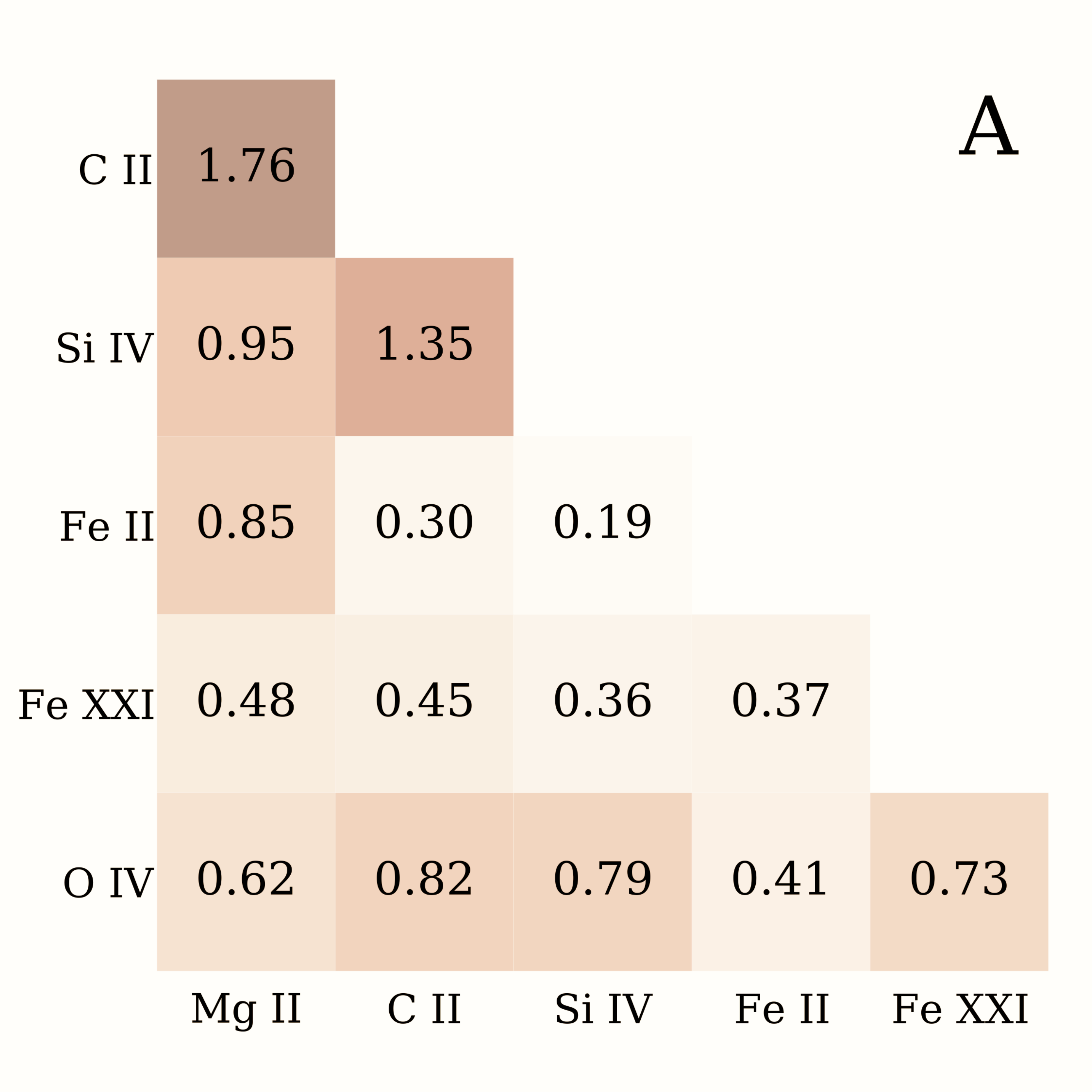}
    \includegraphics[trim={0cm 0cm 0cm 0cm},clip, width=0.32\textwidth]{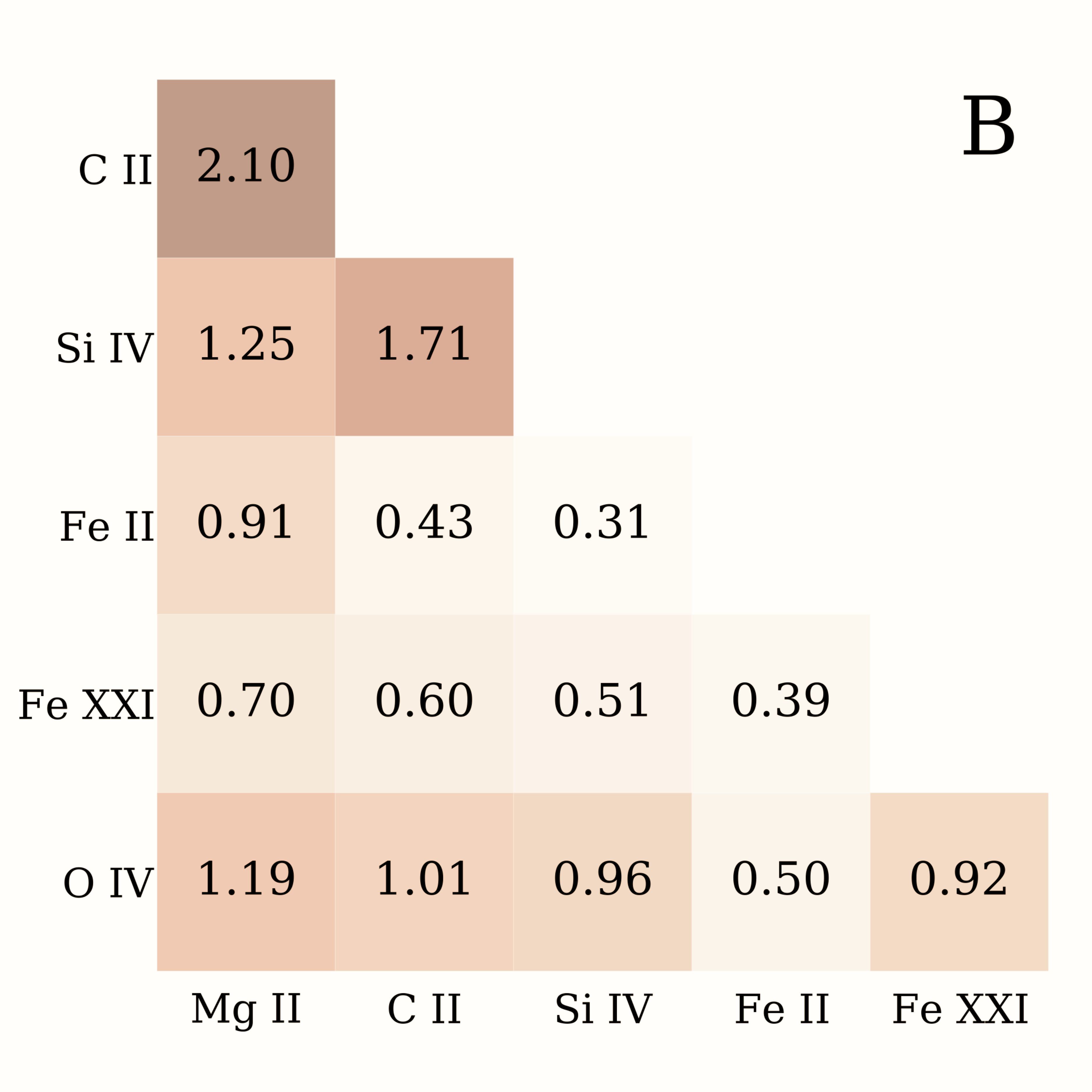}
    \includegraphics[trim={0cm 0cm 0cm 0cm},clip, width=0.32\textwidth]{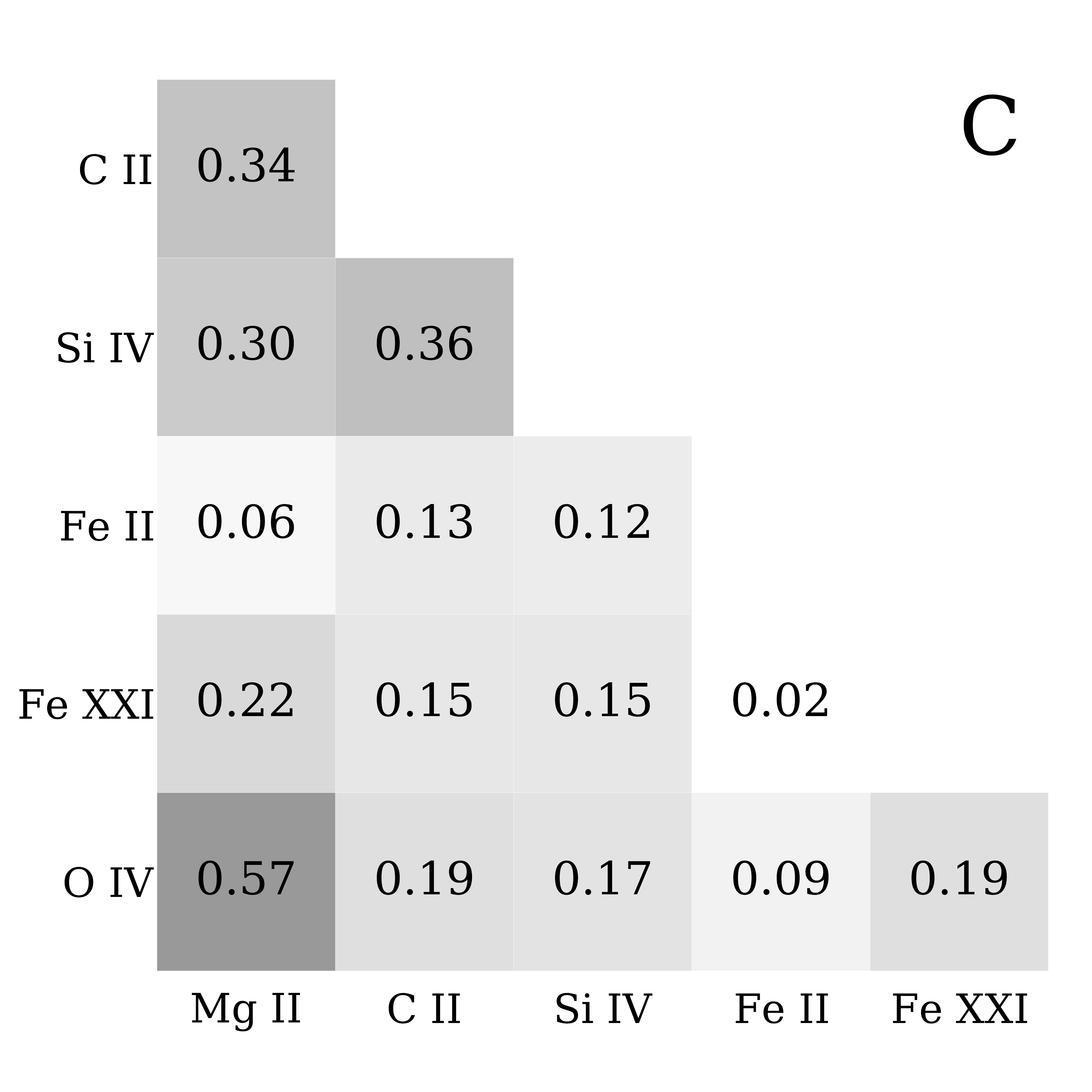} 
\caption{Panel A: MI calculated using the MINE network over both orange and yellow pixels (example in Figure \ref{qs_masks}), i.e., weak flare criterion. Note that unlike the categorical method, the raw values of the MI for the case of the MINE network cannot be normalized into the range $\text{MI}\in[0,1]$. Panel B: the same as panel A but for the stronger flaring criterion, where we only calculate the MI for yellow pixels. These pixels are assumed to coincide predominantly with the flare ribbons. Panel C: difference $(B-A)$. We see that all correlations increase when considering the stricter flare criterion. Line correlations are enhanced over more energetic regions, specifically over the flare ribbons.}
\label{Res_2}
\end{figure*}

It is clear from above that the MI is inhomogeneously distributed over time. To further support the idea that the MI is also unequally distributed spatially, we performed MI calculations with the MINE network over the entire flare data set adhering to two different criteria: (1) A weak flare condition, that includes both the yellow and orange pixels into the calculations as seen for example in Figure \ref{qs_masks}, and (2) q strong flare criteria, where we only calculate the MI for yellow pixels, which are assumed to coincide with the flare ribbons. The results can be seen in Figure \ref{Res_2}. It appears that all line-couplings experiences an increase in correlation when confined to highly energetic portions of the flare. Generally the strongly correlated lines experience a greater enhancement in MI than the weaker correlated lines, except for the case of (\ion{Mg}{2}|\ion{O}{4}), which experiences a significant jump from 0.62 to 1.19. \\

The above analysis can be generalized as follows: If spectral line-pairs do not have high MI scores, the conclusion of absolute independence is unwarranted. For example, it may be possible (such as the above case), that a line-pair with a low aggregate MI, concentrates its correlations within a small subset of spectra, in which case, these spectra would still have very predictable counterparts in their partner line. It is therefore important not only to consider the aggregate MI, but also how this MI is distributed amongst the individual spectra. These considerations segue into the idea of point-wise MI (PMI), a concept that we will address fully in the second paper.

\section{Discussion}
\label{Discussion_section}
The results indicate that for the six spectral lines compared, the correlations are almost entirely suppressed under quiet Sun conditions, which can be seen in panel C of Figure \ref{Res}. The natural conclusion from these results would be that for the quiet Sun, the layers of the solar atmosphere at different scale heights behave largely independently from one another. If there is a communication between the different layers, the rate of energy transfer might be on a time scale that cannot be appreciated by a pixel-wise analysis. Alternatively, since we have not exhausted all possible variables such as intensity, the line shape may not be a suitable observable for capturing the flow of energy through the atmosphere of the quiet Sun.

Panel B  makes it clear that during solar flares, the correlations turn on and become significant for lines with similar formation heights. As expected, the two resonant lines of \ion{Mg}{2} and \ion{C}{2}, which share a similar formation height, also have the highest degree of correlation with a MI of 0.5, followed by (\ion{C}{2}|\ion{Si}{4}), and finally (\ion{Mg}{2}|\ion{Si}{4}). This is also evident in the first three panels of Figure \ref{MIC}. On the other hand, both the \ion{O}{4} and Iron lines are very weakly correlated with any other line. This low correlation may be due in part to a poor signal-to-noise ratios, since all three of the previously mentioned lines are comparatively weaker than the major lines discussed above, and in part due to the finite speed of energy transport between the different formation heights. Additionally, it is surprising that the (\ion{Mg}{2}|\ion{O}{4}) line-pair has a higher or equal correlation to that of the two transition region lines (\ion{Si}{4}|\ion{O}{4}), which are blended at $1401.51~\text{\AA}$ and share a similar formation temperature.

In order to examine how poor signal-to-noise ratios impact the results, we repeated the numerical calculations after removing noisy \ion{Si}{4} and \ion{O}{4} spectra. We found that the couplings for these lines experienced marginal increases; however, the line-pair (\ion{Mg}{2}|\ion{O}{4}) had no less correlation than that of (\ion{Si}{4}|\ion{O}{4}). A possible explanation might be that \ion{Si}{4} emission becomes momentarily optically thick during the impulsive phase, and that its formation height could be compressed into a small region containing both the \ion{Mg}{2} and \ion{C}{2} emission \citep{Kerr_2019}. This drop in formation height would explain why the correlations between \ion{Si}{4} and the chromospheric lines are so high, while simultaneously explaining the identical MIC scores between \ion{O}{4} and (\ion{Mg}{2}, \ion{C}{2}, and \ion{Si}{4}).

To address the possibility of dependencies existing at time lags, we performed an exhaustive study over the temporal domain for all line-pairs and for all 21 flares independently. In this case, the lower baseline estimate of the uncorrelated state for the MINE network was achieved in the same way as before, by sampling $\mathcal{L}1$ and $\mathcal{L}2$ from pixels $A$ and $B$ freely across space and time. However, the upper channel of the network received $\mathcal{L}1$ from pixel $A$ and $\mathcal{L}2$ from pixel $A+n\Delta t$, where $n$ is the number of rasters in the observation, and $\Delta t$ is the cadence. Independent models were then trained for all integers $n$ and the MI followed suit. We found that in general, the MI between different spectral lines steadily decreases with increased time lag; however, for flares 6, 17,  and 20 in Table \ref{obs} the line-pair (\ion{C}{2}|\ion{Fe}{2}) obtained a maximum MI at a consistent time lag of 2.5 minutes. This exceptional subset requires more analysis, but for the most part, it is clear that the MI results obtained in this study are best analyzed instantaneously. We do not present the details of this time series analysis here, but will include it in a future paper. It should however be mentioned that IRIS data are not ideal for this type of analysis, since $\Delta t$ can vary between observations.\\

The design of our analysis allows us to gain confidence with our results. Even though we calculated the same quantities using two very different approaches, both the categorical and numeric methods return highly complementary results as seen in panels A and D of figure \ref{Res}. The magnitudes of the results are not however comparable, since the numerical method does not allow us to estimate the individual entropies of each line, which are necessary to compress the MI into its normalized form over the domain $\text{MI}\in[0,1]$. Regardless of this detail, both methods are almost in a one-to-one agreement with regards to the order of line-pair correlation strengths.\\

The interpretation of low MI scores requires a subtle analysis. Since we now know that the quiet Sun at best only weakly couples the shapes of spectra from different layers of the solar atmosphere, the preprocessing step of quiet Sun masking could play a crucial role in the interpretation of our results. For instance, we  mentioned in section \ref{VAE_sec}, that each line defines the extent of its flaring region differently. That is, \ion{Fe}{2} only has spectral shapes that differ substantially from the spectral shapes of the quiet Sun, directly over the flare ribbon, while on the other hand, \ion{Mg}{2} has a large flaring area that encompasses the flaring region of \ion{Fe}{2} and much more. Therefore, part of the MI score reflects the size of the line-specific flaring region. If instead of defining the flaring region in terms of \ion{Mg}{2},  we chose to define it in terms of \ion{Fe}{2}, then we would expect the correlations between weakly coupled lines to increase.

The analysis of the temporal behavior of the MI between magnesium and the three weaker lines during solar flare 13 in Table~\ref{obs} strengthens this argument. Figure \ref{EV_MI} makes it clear that these lines can achieve MI scores much higher than anticipated from Panel B in Figure \ref{Res}. Since the MI curves in Figure \ref{EV_MI} closely follow the GOES X-ray signatures, and since this particular  observation had ribbons that eventually covered a large portion of the IRIS FOV, we can tentatively conclude that weak lines like \ion{Fe}{2} only couple to different atmospheric heights directly over the flare ribbons, where the signal-to-noise ratio is greatly improved, and violent atmospheric motions knit the entire column together. This conclusion gains more support when comparing the MI scores calculated for both the weak (orange and yellow pixels) and strong (yellow pixels) flare conditions. Since the VAE reliably assigns yellow labels to regions over the flare ribbon, an enhancement in the MI indicates a positive relationship between correlation strength and energy deposition.

Furthermore, even though some spectral lines show a weak aggregate MI score, they may nevertheless contain highly correlated spectra. This would be the case if the majority of the MI (even if it were small) were concentrated among a small subset of spectra.These considerations make it clear that an analysis of the PMI is indispensable.

\section{Conclusions and Outlook}
\label{conclusion_section}

We analyzed the dependencies between six spectral lines in terms of profile shapes (not absolute intensities). The lines selected sample a large range of atmospheric heights from the photosphere up to the transition region, with the goal of finding correlations between the thermodynamics of the different solar atmospheric layers. We looked at an information-theoretic measure, called MI, which captures both linear and nonlinear dependencies between spectra. We used two different approaches to calculate the MI: a categorical binning approach with the k-means algorithm, and a numerical approach with a NN architecture called MINE.\\

We obtained the following list of conclusions.
\begin{itemize}
\item Both approaches returned complementary results and therefore support the accuracy of our calculations. 
\item The numerical approach is preferable over the categorical approach for several reasons. It is automatic and robust, allowing us to work with the raw spectral data directly. The results are largely independent of the input dimensions and are insensitive to the specifics of the network architecture, with different architectures and more complex layers simply resulting in shorter convergence times. This implies that unlike the categorical approach, the MINE network does not introduce any hyperparametric dependencies, and leaves us with accurate objective estimations of the true MI.
\item Under quiet Sun conditions, the lines are almost entirely decoupled, which implies one of two conclusions: (1) The different height layers of the solar atmosphere operate independently from one another and are largely uncoupled. This uncoupling is only in terms of line shape, and we do not rule out the possibility of intensity correlations. (2) The time scale for energy transfer between the different atmospheric layers, absent of small-scale activity, is so gradual in comparison to the local effects, that any coupling in terms of line shape is washed out before the energy deposition reaches the area of effect, and is therefore impossible to untangle.
\item The correlations between line shapes turn on during solar flares. There appears to be two groups of lines, weakly and strongly coupling lines. The latter include \ion{Mg}{2}, \ion{Si}{4} and \ion{C}{2}, which show significantly high MI scores, with the highest being between the (\ion{Mg}{2}|\ion{C}{2})  line-pair, with a score of 0.5. Since these lines are both optically thick resonant lines that share a similar formation height, this result bolsters the confidence of our methods.
\item The remaining spectral lines: \ion{Fe}{2}, \ion{Fe}{21}, and \ion{O}{4} have significantly weaker coupling capacities. These couplings are only weak on aggregate. A single flare study with a large unfolding double ribbon indicates that the strength of the couplings is closely correlated to the GOES soft X-ray derivative, and becomes significantly enhanced over the flare ribbons.
\item The MI between all line-pairs is enhanced when considering spectra only sourced from pixels over the flare ribbon.
\end{itemize}

In conclusion, there appears to be a direct relationship between energy deposition and the strength of correlations between line-pairs. We initially observed that  the coupling strengths under quiet Sun conditions were negligible. These couplings became significant during solar flares. The correlations were further enhanced during the impulsive phase of flares and when considering only spectra from pixels associated with the flare ribbons. This allows us to conclude that the solar atmosphere is coupled from the photosphere to the corona in regions of strong energy deposition during large M- and X-class solar flares.\\

Since we now know that the MI is amorphously distributed both in time (impulsive phase) and space (flare ribbons), it becomes crucial to investigate how the aggregate MI is distributed amongst individual spectra. This quantity is known as the PMI, and along with conditional probability distributions, is the subject of the next paper. By knowing the PMI, the coupling of the different solar layers can be investigated in more detail, and it becomes possible to predict spectral line shapes and their occurrence from one line to another. The methods used in this paper can easily be extended and applied to the analysis of smaller flares, which would greatly extend the statistical validity of these results. It would also be of interest to investigate whether these correlations vary between smaller and larger flares, since there may exist fundamental differences in ionization equilibrium timescales \citep{Kerr_mgII_2}. Additionally, the MINE network represents a highly versatile tool, which could also be used to examine the mutual information between two different data types, such as images and spectra.

\acknowledgements
We would like to thank Säm Krucker, Cédric Huwyler, Martin Melchior, and Denis Ullmann for their fruitful discussions. We used the Scikit-Learn module for implementations of the k-means algorithm \citep{SK}, and the syntactically python-friendly high-performance deep learning framework PyTorch, for the construction of the MINE network \citep{PyTorch}. The preprocessing was done in IRISreader, a library specifically developed for handling large volumes of IRIS data \citep{IRISreader}. We would like to thank the Swiss National Science Foundation for funding this research under grant No. 407540\_167158, as well as LMSAL and NASA for allowing us to download all of the IRIS data from their servers. IRIS is a NASA small explorer mission developed and operated by LMSAL with mission operations executed at NASA Ames Research center and major contributions to downlink communications funded by ESA and the Norwegian Space Centre.

\bibliographystyle{apj}
\bibliography{journals,references}

\end{document}